\documentclass{siamart190516}

\usepackage[utf8]{inputenc}
\usepackage{graphicx}
\usepackage{lipsum}% http://ctan.org/pkg/lipsum
\usepackage{multirow}
\usepackage{graphicx}% Include figure files
\usepackage{dcolumn}% Align table columns on decimal point
\usepackage{bm}% bold math
\usepackage{float}
\usepackage[T1]{fontenc}
\usepackage{mathptmx}
\usepackage{mathtools}
\DeclareMathAlphabet{\mathcal}{OMS}{cmsy}{m}{n}

\usepackage{amsmath, amssymb}
\usepackage{xcolor}
\usepackage{mathrsfs}
\usepackage{booktabs}
\usepackage{hyperref}
\usepackage{caption}
\usepackage{subcaption}
\usepackage{changepage}   % for the adjustwidth environment
\usepackage[new]{old-arrows}
\usepackage{wrapfig}
\usepackage{color}
\usepackage{tikz-cd}
\usepackage{fullpage}
\usepackage{amssymb}
\usepackage{,amsmath}
\usepackage{tikz}
\usepackage{ntheorem}
\usepackage{array}
\usepackage{tikz-cd}

\newcommand{\blockpar}{\parindent 0pt \parskip 5pt}

\blockpar
\newcommand{\func}[3]{#1 \colon #2 \rightarrow #3}	% Function definition
\newcommand{\fig}[1]{Fig.~\ref{#1}}

\newcommand{\ints}{\mathbb{Z}}

\newcommand{\sub}{\subseteq}		% Subset
\newcommand{\define}{ \mathrel{\bf \colon \kern -2pt =} }	% Definition

\renewcommand{\int}{\cap}
\renewcommand{\(}[1]{\begin{equation} \label{#1}}
\renewcommand{\)}{\end{equation}}
 \usepackage{subcaption}
\newdimen\figrasterwd
\figrasterwd\textwidth
\usepackage{cite}

\newcommand{\multilinecomment}[1]{}

\title{Understanding High-Order Network Structure using Permissible Walks on Attributed Hypergraphs}

\author{
Enzo Battistella\thanks{Network Science Institute, Northeastern University, Boston, USA.} 
\and Sean English\thanks{University of North Carolina Wilmington, \texttt{EnglishS@uncw.edu}} 
\and Robert Green\thanks{Pacific Northwest National Laboratory and State University of New York at Albany, Albany, New York} 
\and Cliff Joslyn\thanks{Pacific Northwest National Laboratory and Binghamton University.} 
\and Evgeniya Lagoda\thanks{Freie Universit\"at Berlin and Berlin Mathematical School} 
\and Van Magnan\thanks{University of Montana} 
\and Audun Myers\thanks{Pacific Northwest National Laboratory.} 
\and Evan D. Nash\thanks{Center for Naval Analyses.} 
\and Michael Robinson\thanks{American University, Washington, D.C.}
}

\date{April, 2024}

\begin{document}

\maketitle

%!TEX root = ..\main.tex
%-------------------------------------------------
%*************************************************

\begin{abstract}
Hypergraphs have been a recent focus of study in mathematical data science as a tool to understand complex networks with high-order connections. One question of particular relevance is how to leverage information carried in hypergraph attributions when doing walk-based techniques. In this work, we focus on a new generalization of a walk in a network that recovers previous approaches and allows for a description of permissible walks in hypergraphs. Permissible walk graphs are constructed by intersecting the attributed $s$-line graph of a hypergraph with a relation respecting graph. The attribution of the hypergraph’s line graph commonly carries over information from categorical and temporal attributions of the original hypergraph. To demonstrate this approach on a temporally attributed example, we apply our framework to a Reddit data set composed of hyperedges as threads and authors as nodes where post times are tracked.
\end{abstract}

% REQUIRED
\begin{keywords}
Network Analysis, Attributed Hypergraphs,  Higher Order Networks, Dynamic Graphs, Walks on Graphs, Information flow, Influence in Networks.
\end{keywords}

% REQUIRED
\begin{AMS}
05C65, 05C76, 05C20 
\end{AMS}
% 	05C65  	Hypergraphs
%	05C21  	Flows in graphs
%   05C76  	Graph operations (line graphs, products, etc.)

%!TEX root = ..\main.tex
%-------------------------------------------------
%*************************************************

\section{Introduction} \label{sec:introduction}

Graph theory has a rich history in the analysis of complex systems as network models with a set of entities (nodes) with pairwise relationships (edges). Graph models benefit from their simplicity and swath of supporting theory. However, many real-world systems are not accurately modeled with just a dyadic connection between nodes, as in a graph data structure.
Graphs naturally generalize to structures known as hypergraphs, in which a set of $k$ nodes that are interacting can instead be modeled by a size $k$ hyperedge, for $k \ge 1$. However, it is notably difficult to generalize many graph-theoretic notions and techniques, for example spectral analysis~\cite{hayashi2020hypergraph,banerjee2021spectrum, cooper2012spectra} and random walks~\cite{chitra2019random}),  to the world of hypergraphs. Instead, a common approach to study hypergraphs is by using other graphs which approximate them to capture the higher-order connections between hyperedges.

Additionally, it is common that both graphs and hypergraphs carry data, called attributes, in the form of numerical or vector weights, or other data types such as Boolean, categorical, or ordinal data. Such attributes can be assigned to vertices, edges, hyperedges, or some combination of these, depending on the application. They can arise from measured data, or can be calculated from other data sources or from inherent properties of the graph itself.

In this work we present a general methodology for approaching walks in hypergraphs that respects a relation on its  attributes. In this method, we create a new graph representation of the hypergraph that summarizes the structure of the underlying hypergraph while respecting the attributes. This approach is able to capture both the connectivity of the hypergraph as well as how the attributed information is related. 

An area of likely use for such a graph structure would be in the study of its traversals, which is a classical area of interest in graph theory. One type of traversal is a \textit{walk}. Given some vertex $u$ in a graph, we can ``walk" from $u$ to another vertex $v$ if $u$ and $v$ are connected by an edge. A walk is then a sequence of vertices $v_1v_2\ldots v_k$ such that each $v_i$ and $v_{i+1}$ are connected by an edge. These walks are used in many applications with a major one being in vectorization of nodes (e.g., node2vec~\cite{grover2016node2vec}), in which data scientists can study the relations between nodes with applications such as classification through clustering and training natural language processing models.

Graph walks are readily understood in application, but they rely on the structure of graphs in such a way that needs to be extended to the language of hypergraphs carefully~\cite{carletti2020random,carletti2021random,chitra2019random}. In a graph, if two vertices are connected by an edge, they are connected by a unique edge (considering only a simple graph and not a multi-graph). So, when a graph walk moves from node $u$ to node $v$, we implicitly understand the edge through which we traverse. Indeed, describing a graph walk in terms of a sequence of vertices, a sequence of edges, or an alternating sequence and vertices and edges, are all equivalent.
But in a hypergraph, two vertices can both be contained in any number of edges. 
Recently in the literature there has been a keen interest in defining hypergraph walks in ways that can be recovered by a  graph structure while still incorporating information from the hypergraph. One such approach are $s$-walks~\cite{aksoy2020hypernetwork}, where an $s$-line graph (see Definition~\ref{def:s_line_graph}) is constructed from the hypergraph. 

There has also been an active community focused on walks in graphs that incorporate temporal information~\cite{starnini2012random,nguyen2018dynamic} as many graphs of interest may evolve over time with edges and nodes having temporal attributes. These call for the development of a mathematical framework of dynamic network science known as temporal networks~\cite{HOLME201297}. Such structures arise in a variety of applied settings, be it social network analysis~\cite{Skyrms2000}, transit network analysis~\cite{DavidBoyce2012}, biological interaction networks such as disease spread~\cite{Husein2019}, or communication networks for space satellites\cite{bernardoni2023algebraic}. 

In this work we study walks in attributed hypergraphs by defining the permissible walk graph of a hypergraph and use it to generalize $s$-walks to attributed hypergraphs. 
In doing so we transfer rich attribute information from the hypergraph to the $s$-line graph structure that would normally be lost in the $s$-walks framework.
This allows for more accurate modeling of data and the resulting application of the corresponding random walks (e.g., node and hyperedge embedding).
Some common attributes that can be found on hypergraphs are temporal and type attributes, where, for example, a hyperedge may only be active during a specific interval of time or the edge may be of a specific type that can only communicate to another hyperedge of that type. The goal of this work is to incorporate this information into the $s$-line graph and show how this better models complex hypergraph data. We demonstrate this through a worked example with temporal attributes.

\paragraph{Related Work}

In \cite{ostroski2021dynamic,ostroski2021dynamic2, ostroski2023scalable} a method developed on the line-graph of a dynamic multi-graph was used for edge clustering. This is similar to our method as it takes a dynamic multi-graph, which is an attributed hypergraph with direction and temporal attributes, and creates an attributed line-graph. However, our method is a generalization of this approach as it can take higher-order networks as well as multiple attribute types. We provide details on how our method generalizes this approach in the Appendix Section~\ref{app:ostroski}
In ~\cite{failla2023attributed} a similar procedure to our initial exhibition in~\cite{battistella2023permissible} was developed in which they take a higher-order network (hypergraph) and construct an attributed stream graph. Our work here again is a generalization of this method based on their modeling of only temporal attributes of higher order interactions. 

\begin{figure}[h!] 
    \centering
    \hspace{-10mm}
    \begin{subfigure}[b]{.6\textwidth}
        \centering
        \includegraphics[width=.8\textwidth]{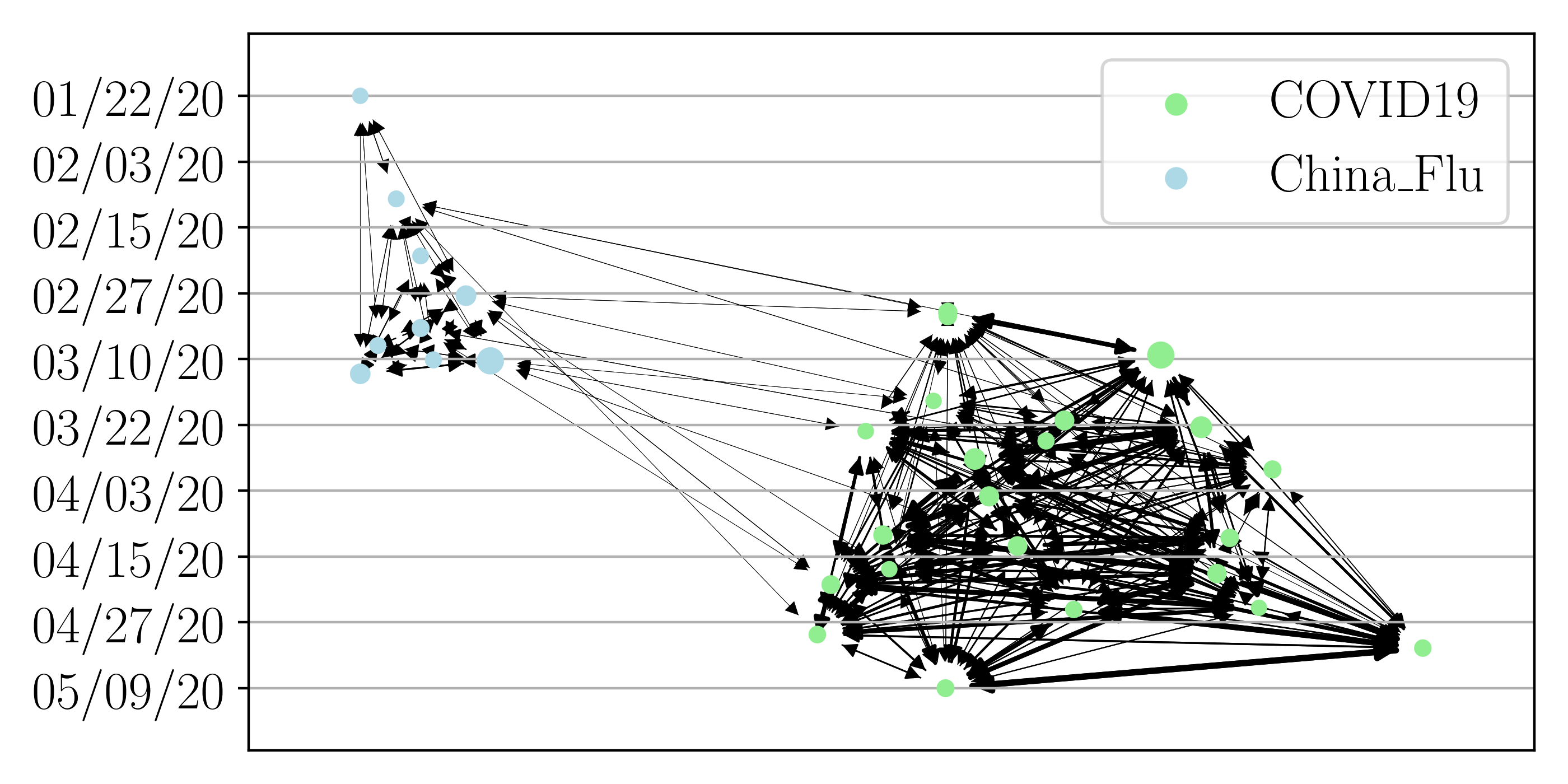}
        \caption{$s$-line graph.}
        \label{fig:intro_figure_line_graph}
    \end{subfigure}
    %\hfill
    \hspace{-5mm}
    \begin{subfigure}[b]{.3\textwidth}
        \centering
        \includegraphics[width=.65\textwidth]{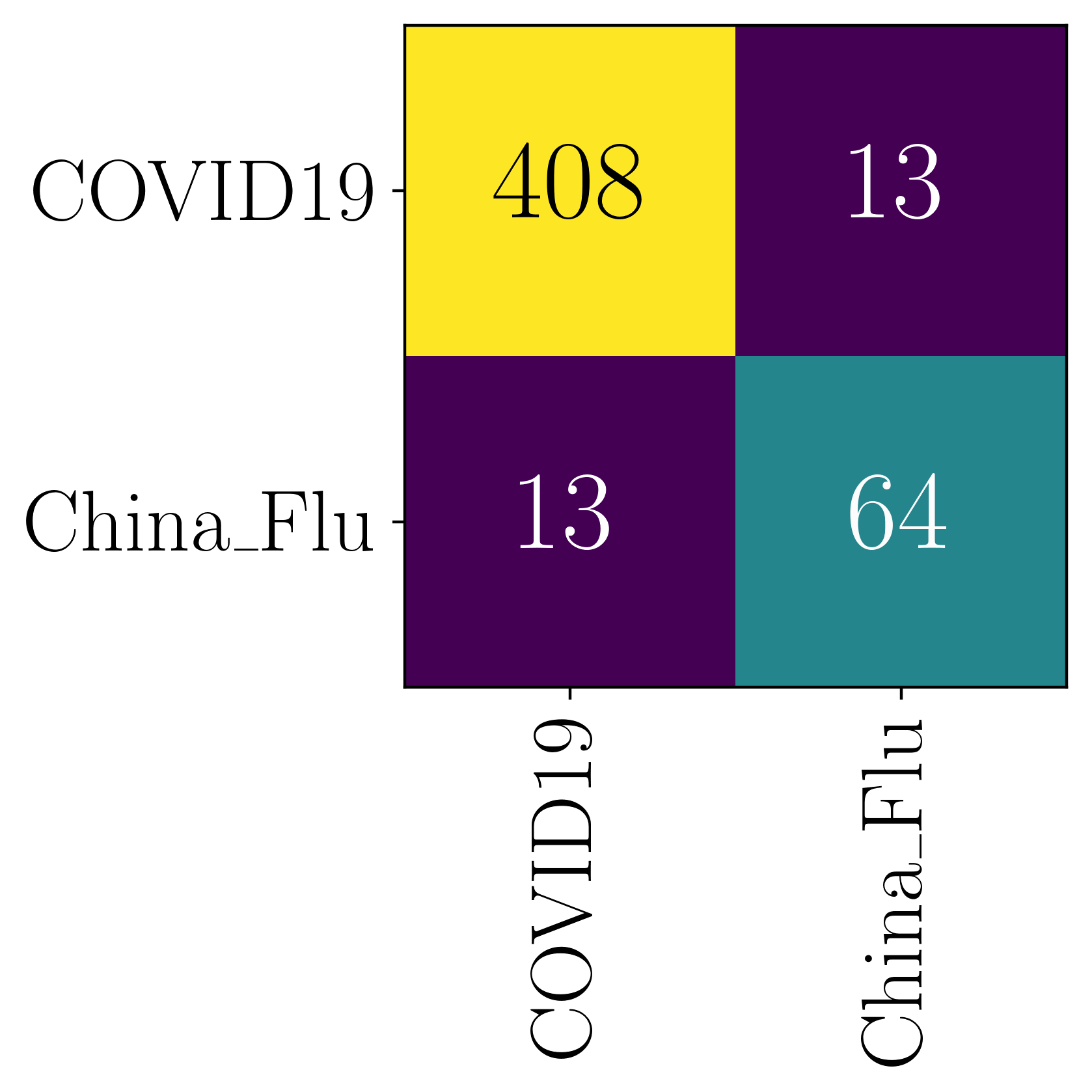}%
        \caption{Interaction matrix for the $s$-line graph.}
        \label{fig:intro_figure_s_line_IM}
    \end{subfigure}
    
    \hspace{-10mm}
    \begin{subfigure}[b]{.6\textwidth}
        \centering
        \includegraphics[width=.8\textwidth]{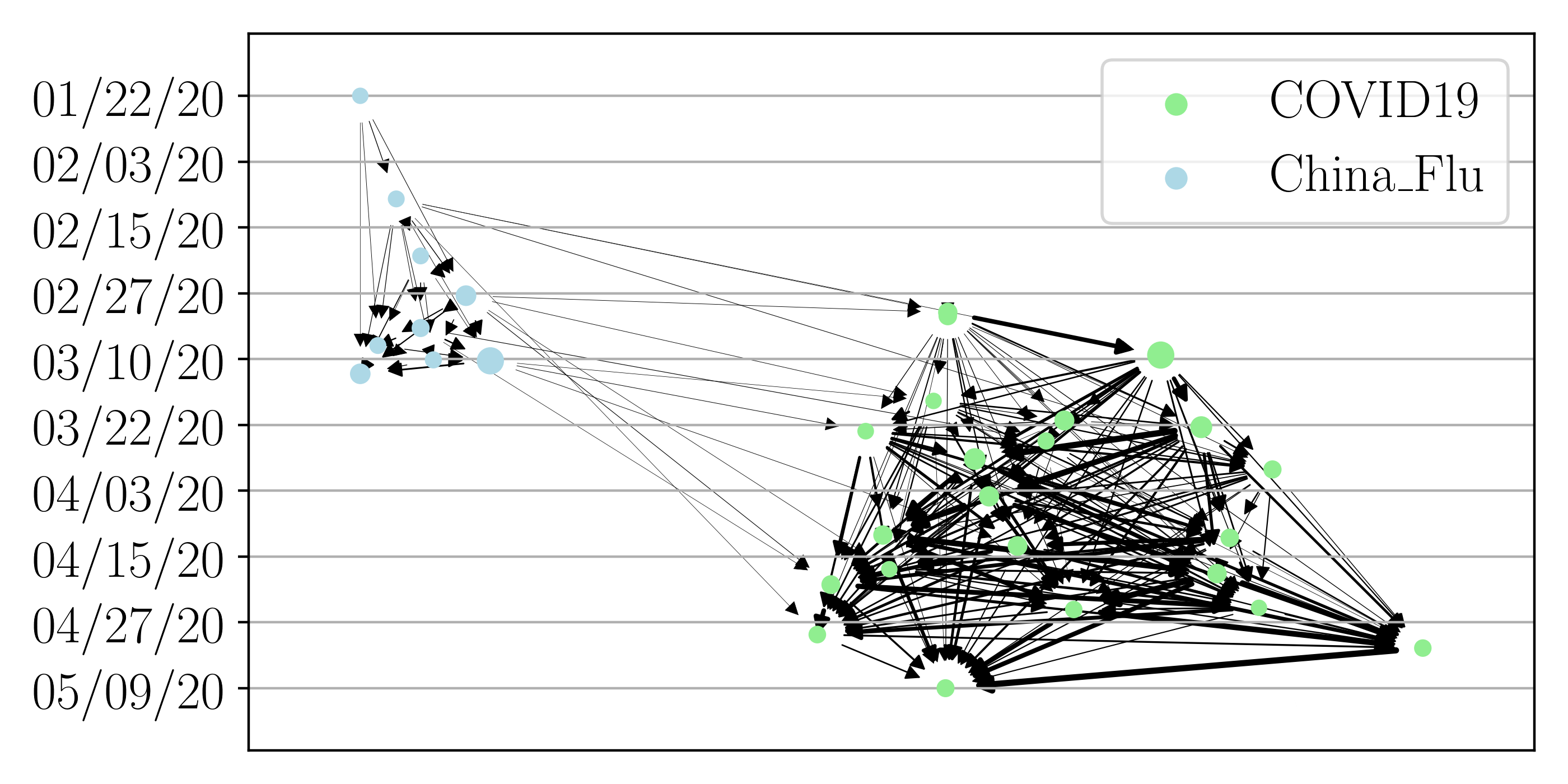}
        \caption{Permissible walk graph.}
        \label{fig:intro_figure_permissible_walk_graph}
    \end{subfigure}
    %\hfill
    \hspace{-5mm}
    \begin{subfigure}[b]{.3\textwidth}
        \centering
        \includegraphics[width=.65\textwidth]{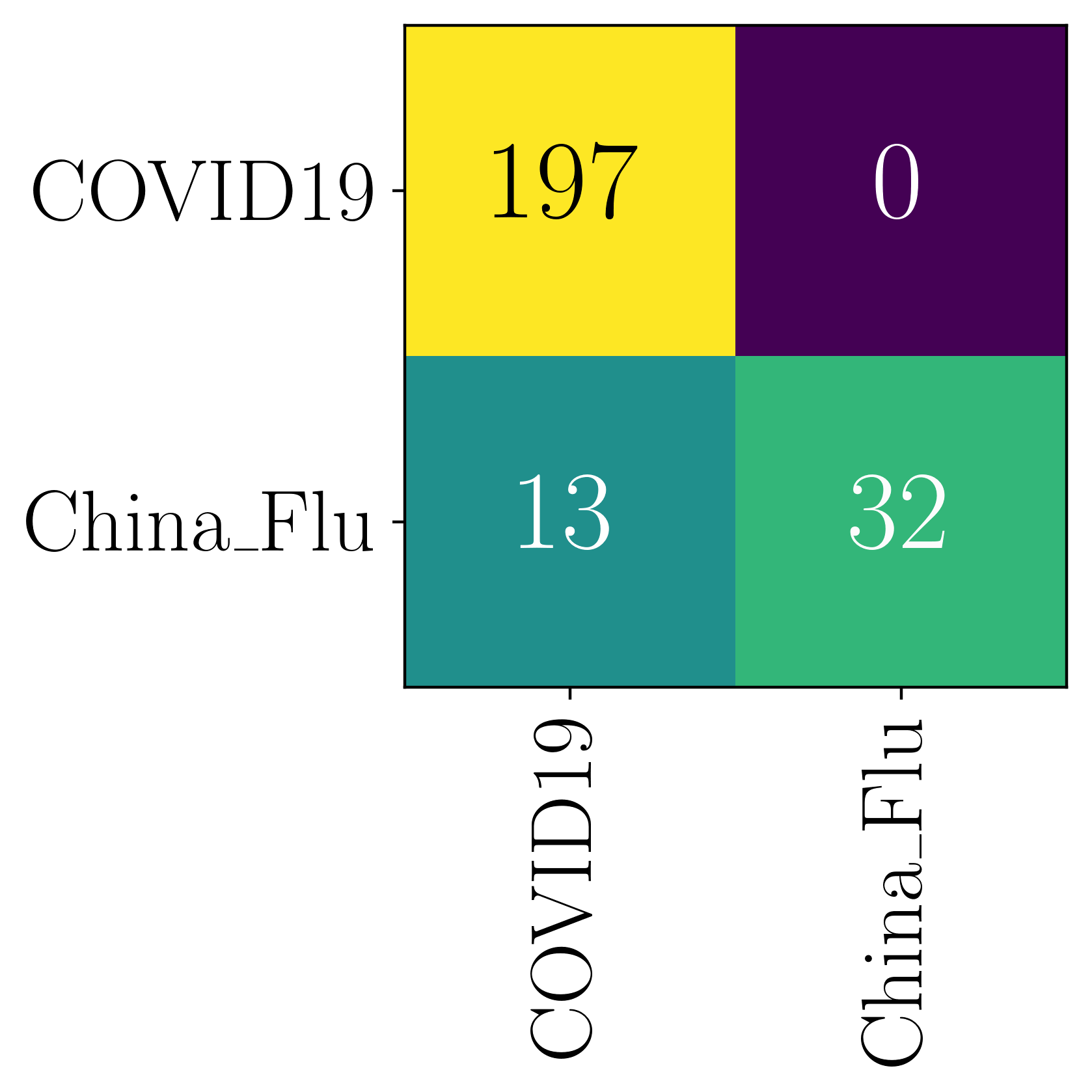}%
        \caption{Interaction matrix for the permissible walk graph.}
        \label{fig:intro_figure_PW_IM}
    \end{subfigure}
    \caption{Comparison between the interaction matrix of the $s$-line graph for $s=10$ in (b) and permissible walk graph in (c) from the attributed hypergraph associated to the China\_Flu and COVID19 subreddits of Reddit.}
    \label{fig:intro_IM_comp_sline_PW}
\end{figure}

\paragraph{Example in Reddit}
To give a glimpse into the importance of considering attributes, in Fig.~\ref{fig:intro_IM_comp_sline_PW} we show both the $s$-line graph and our permissible walk graph of a temporally attributed hypergraph modeling Reddit communication between two subreddits: China\_Flu and COVID19. 
In Fig.~\ref{fig:intro_figure_line_graph} we show the $s$-line graph of the underlying hypergraph for $s=10$ where the nodes in the graph represent the hyperedges of the underlying hypergraph where hyperedges are threads that are composed of sets of authors (only threads with at least 400 authors are used for visualization purposes). Each edge in this graph is bi-directional as the temporal attributes are not considered in this model.
In comparison, in Fig.~\ref{fig:intro_figure_permissible_walk_graph} we demonstrate the permissible walk graph of the same attributed hypergraph but now the edges follow the temporal order.
Specifically, we add directed edges between nodes (threads) if a thread was active before another thread. By doing this we are able to capture the flow of thread creation and subreddit creation that would normally be lost when just considering the $s$-line graph connectivity. 
We quantify this as shown in Fig.~\ref{fig:intro_figure_s_line_IM} and \ref{fig:intro_figure_PW_IM}, where we count the number of edges between the subreddits and represent it as a matrix. Fig.~\ref{fig:intro_figure_s_line_IM} is for the $s$-line graph where the interaction is symmetric  since each edge is bidirectional. This graph loses the temporal flow of thread and subreddit creation.
In comparison, Fig.~\ref{fig:intro_figure_PW_IM}  shows this interaction matrix for the permissible walk graph which is asymmetric with more edges going from China\_Flu to COVID19 capturing the temporal directionality of the subreddit creation and activity.

This example demonstrates the importance of incorporating attribute information in the permissible walk graph of an attributed hypergraph when tracking the evolution of a temporal hypergraph. Additionally, it provides more information on the flow of conversations between these subreddits that is normally lost when using the $s$-line graph.

\paragraph{Organization}
This work begins by introducing the permissible walk graph in Section~\ref{sec:method} and explaining how it is formed from an attributed hypergraph. In Section~\ref{ssec:implementation} we overview many of the details for implementing the permissible walk graph for studying attributed hypergraphs, including two examples illustrating possible relations defined on the attributes.
In Section~\ref{sec:results} we use the permissible walk graph to study a Reddit social network alluded to in Fig.~\ref{fig:intro_IM_comp_sline_PW} where we in detail describe some of the benefits of studying such a dataset from the permissible walk framework.
Lastly, in Section~\ref{sec:conclusions} we provide concluding remarks and future research for permissible walk graphs.
%!TEX root = ..\main.tex
%-------------------------------------------------
%*************************************************

\section{Method} \label{sec:method}

\begin{figure}[h!]
    \centering
    \includegraphics[width=.8\textwidth]{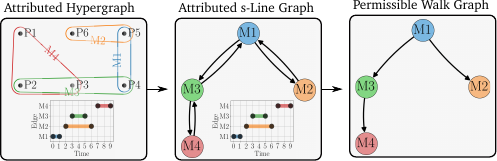}%
    \caption{The pipeline for generating the permissible walk graph starting with an attributed hypergraph, which is used to construct the attributed $s$-line graph using an attribute transferring function. The permissible walk graph is then constructed as an attribute respecting subgraph of the $s$-line graph.}
    \label{fig:pipeline_permissible_walks}
\end{figure}

In this section we present our methodology for constructing a permissible walk graph from an attributed hypergraph. The general procedure is {illustrated} in Fig.~\ref{fig:pipeline_permissible_walks} going from an attributed hypergraph to an attributed $s$-line graph, and lastly the permissible walk graph. Following this section we introduce a small example in Section~\ref{ssec:implementation} to better demonstrate the implementation of the permissible walk graph for studying an attributed hypergraph.

Following Fig.~\ref{fig:pipeline_permissible_walks}, we begin by defining an attributed hypergraph.
\begin{definition}
Let $H = (V, E, \phi, \epsilon, \gamma)$ be an {\bf attributed hypergraph}, where:

\begin{itemize}
    
\item $V = \{ v_i \}_{i=1}^n$ is a finite, non-empty set of {\bf vertices};
\item $E = \{ e_j \}_{j=1}^m$ is a finite, non-empty set of {\bf edges} or {\bf hyperedges} $e_j \sub V$; 
\item $\func{\phi}{V}{X}$ is the {\bf vertex attribute function} to a set of vertex attributes $X$;    
%\item Let We use $I=V \times E$ the {\bf incidence matrix}, such that each {\bf incidence} $I[i][j]$ is 1 if  
\item $\func{\epsilon}{E}{Y}$ is the {\bf edge attribute function} to a set of edge attributes $Y$; and     
\item $\func{\gamma}{I}{Z}$ is the {\bf incidence attribute function}, where $I\subseteq V\times E$ is the collection of all incidences (i.e. $(v,e)\in R$ if $v\in e$, see discussion below), and $Z$ is a set of incidence attributes.

\end{itemize}
\end{definition}

To explain the use of the term incidence attribute function above, every hypergraph $H$ yields a binary relation $I \sub V \times E$ between the vertex set $V$ and the edge set $E$ where if $v\in e$ then $vIe$, which is the domain of $\gamma$. One could also define a Boolean {\bf incidence function} $I': V \times E \to \{0,1\}$, where $I'(v,e)=1$ if and only if $v\in e$, or a Boolean  \textbf{incidence matrix} $I''$, where rows represents elements of $V$, columns represent elements of $E$ and $I''[v_i][e_j] = 1$ if and only if $v_i \in e_j$. In a slight abuse of notation, we will use the symbol $I$ for all three of the incidence relation, function and matrix. The pair  $(v_i,e_j)\in I$ is called an individual {\bf incidence}. %, meaning that the vertex $v_i$ is a member of the edge $e_j$.
%The incidence attribute function $\gamma$ is thus a function on the support of the Boolean function $I$ (i.e., $\gamma(v_i,e_j) \neq \emptyset$ if $I(v_i,e_j) = 1$). 
We call the vertices $v_i\in V$ and the edges $e_j \in E$ generically \textbf{objects}. \footnote{ Note that our definitions also operate for the dual hypergraph.
Viewing $I$ as the incidence matrix we get a natural way to define the \textbf{dual} hypergraph $H^*$ which has $E$ as its vertex set $V$ as its edge set and $I^T$ as its incidence matrix, if $H$ is attributed then the attributes carry to the dual in a natural way with $H^* = (E, V, \epsilon, \phi, \gamma^* )$ where we let $I^*\subseteq E\times V$ be the subset such that $(e,v)\in I^*$ if $e\ni v$, and $\func{\gamma^*}{I^*}{Z}$, where $\gamma^*(e_i,v_j) = \gamma(v_i,e_j)$. 
When considering duals we call the original $H$ the \textbf{primal} and $H^*$ the dual, noting that this is arbitrary. For the remainder of this work we operate in the primal pointing of the attributed hypergraphs.}

%Treating $I$ as an incidence function we can naturally define the \caj{\textbf{marginal} image} functions of $H$ as $I_V : V \to 2^E$ and $I_E : \to 2^E$ with respective definitions
%\[  I_V(v_i) = \{e_j \in E \mid I(v_i,e_j) = 1\} \qquad I_E(e_j) = \{v_i \in V \mid I(v_i,e_j) = 1\}  \]

Following the procedure in Fig.~\ref{fig:pipeline_permissible_walks} we next  construct the attributed $s$-line graph by first building the underlying $s$-line graph.
\begin{definition}
\label{def:s_line_graph}
For $s\in \mathbb{Z}_{\geq 0}$, the \textbf{$s$-line graph} $L_s = (E, \hat{K})$ of a hypergraph $H = (V, E)$ is a simple graph constructed as follows:
\begin{enumerate}
  \item The node set $E$ of $L_s$ is the set of hyperedges $E$ of $H$.
  \item $\hat{K} \sub \binom{E}{2}$ is a set of undirected graph edges between the vertices $e_i \in E$, where for each pair of hyperedges $e_i \neq e_{i'} \in E$, there exists an edge $\{e_i, e_{i'}\} \in \hat{K}$ if and only if ${|e_i \cap e_{i'}| \geq s}$.
\end{enumerate}
\end{definition}

\begin{figure}[h!] 
    \centering
    \begin{subfigure}[b]{.2\textwidth}
        \centering
        \includegraphics[width=.99\textwidth]{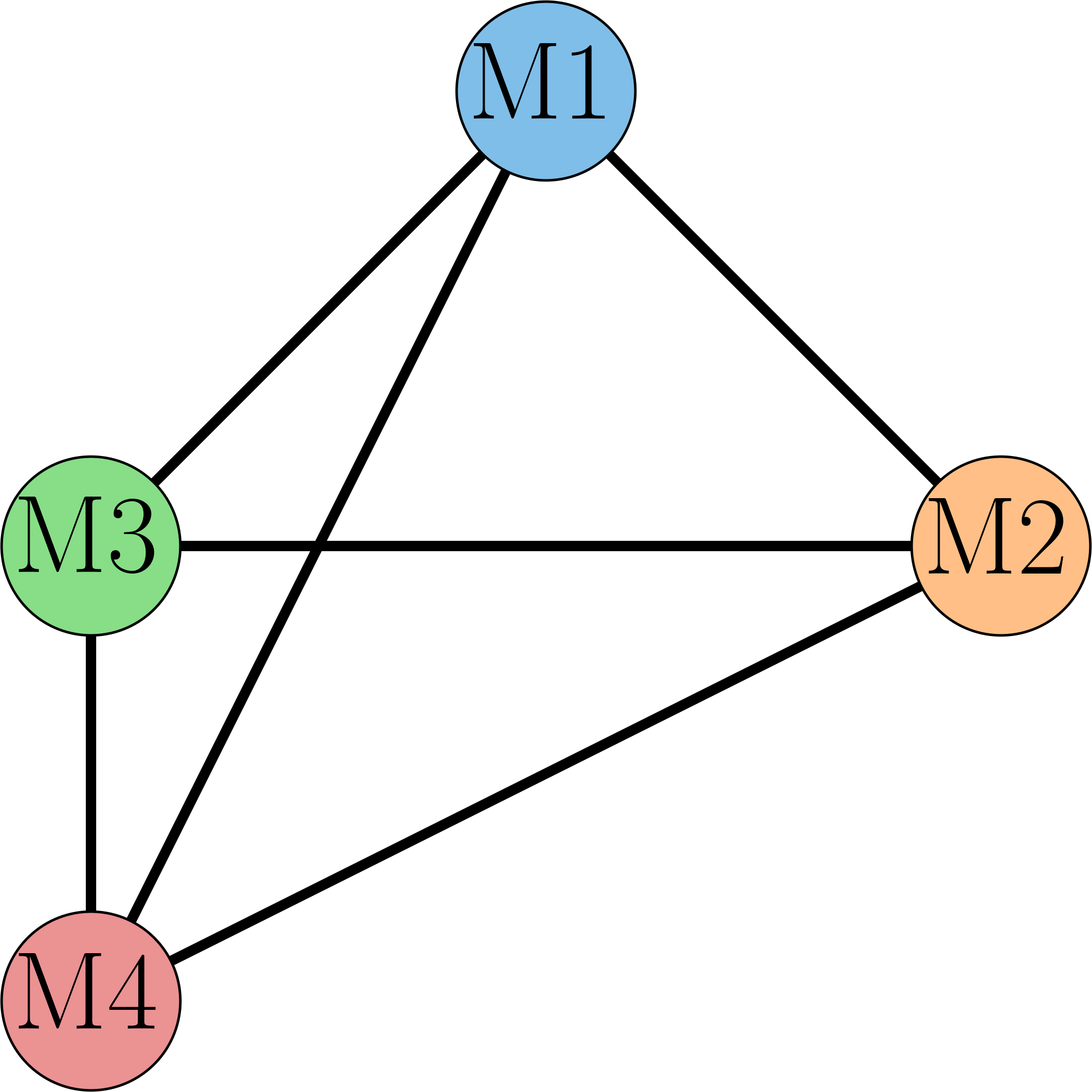}
        \caption{$0$-line graph.}
        \label{fig:meeting_0_line_graph}
    \end{subfigure}
    \hfill
    \begin{subfigure}[b]{.2\textwidth}
        \centering
        \includegraphics[width=.99\textwidth]{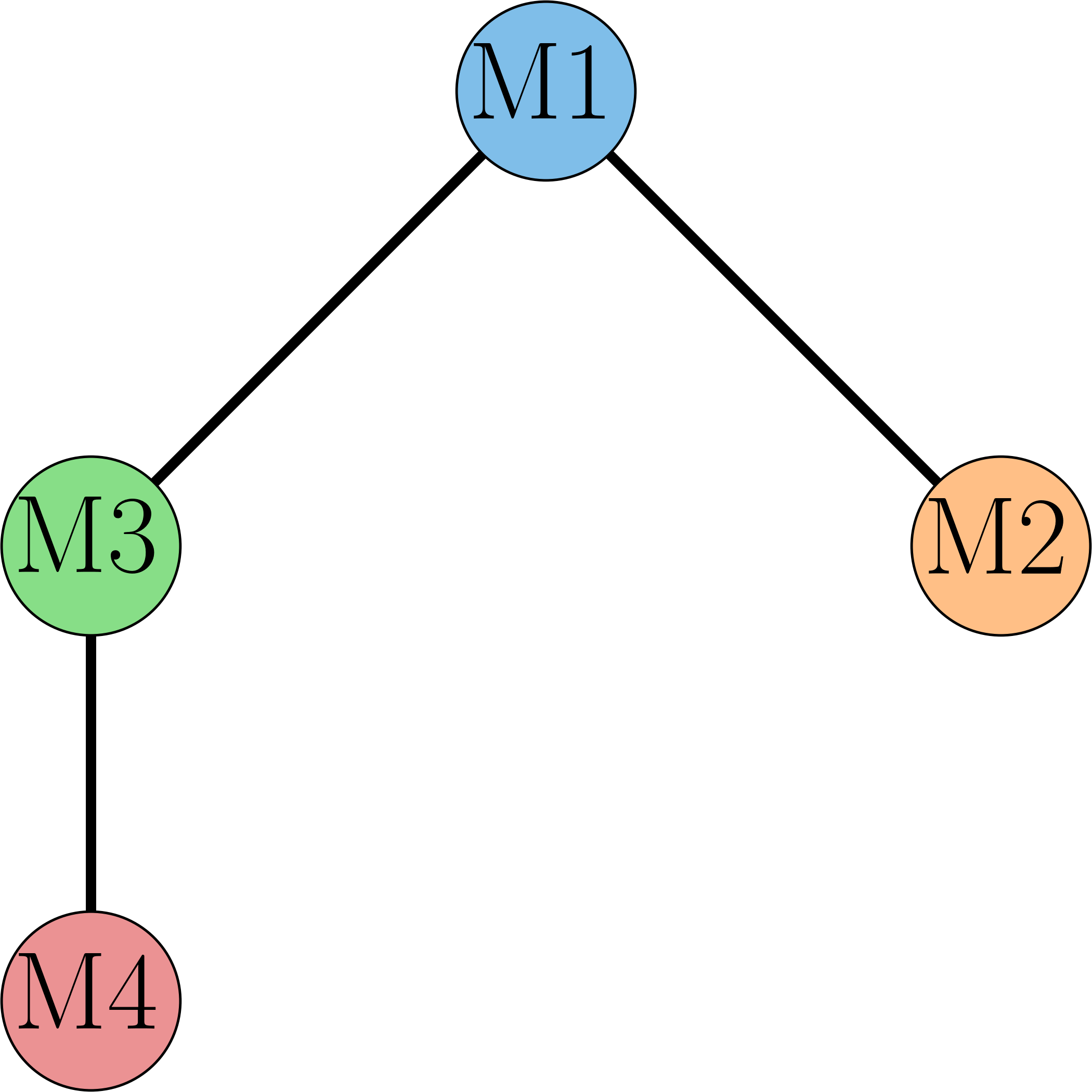}
        \caption{$1$-line graph.}
        \label{fig:meeting_1_line_graph}
    \end{subfigure}
    \hfill
    \begin{subfigure}[b]{.2\textwidth}
        \centering
        \includegraphics[width=.99\textwidth]{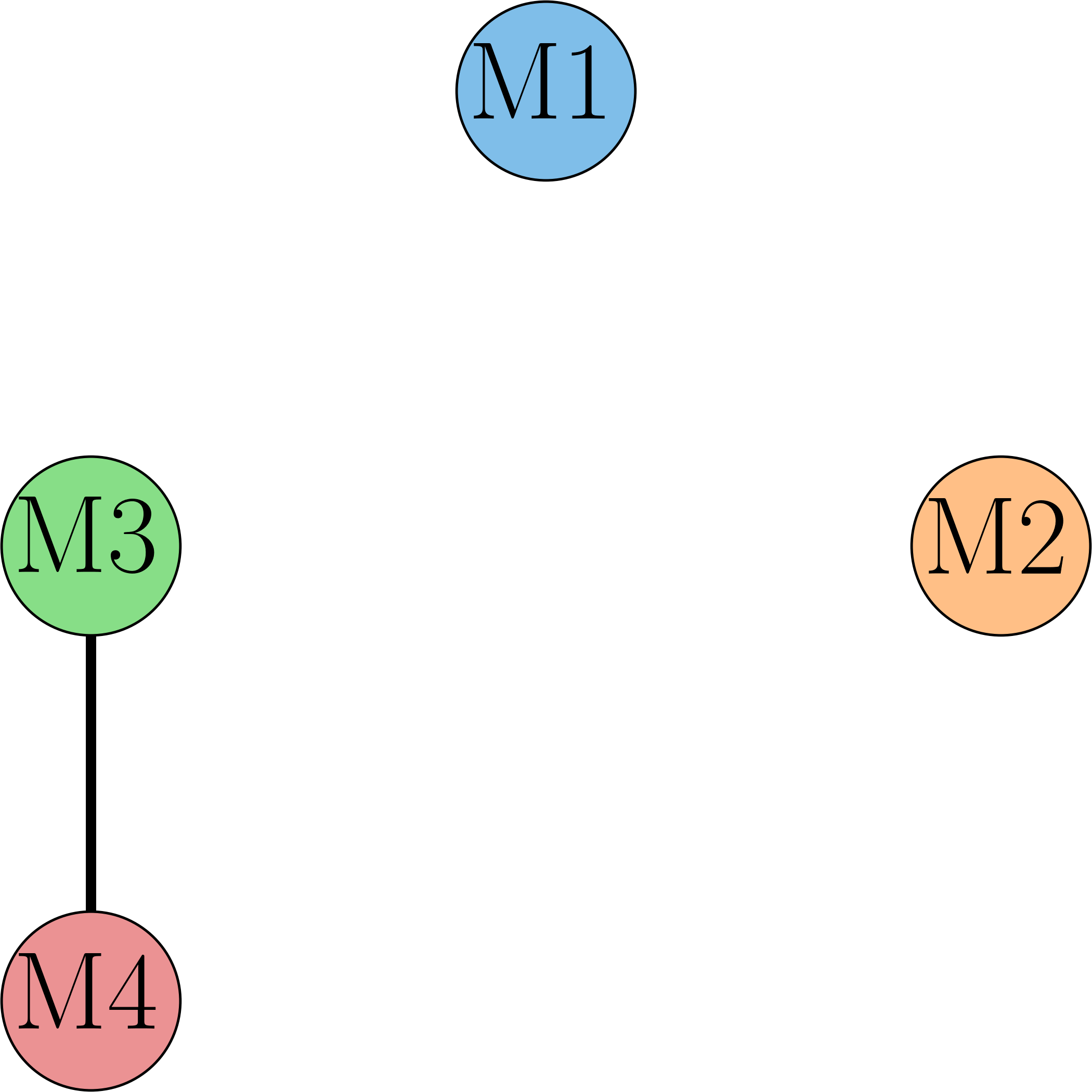}
        \caption{$2$-line graph.}
        \label{fig:meeting_2_line_graph}
    \end{subfigure}
    \hfill
    \begin{subfigure}[b]{.2\textwidth}
        \centering
        \includegraphics[width=.99\textwidth]{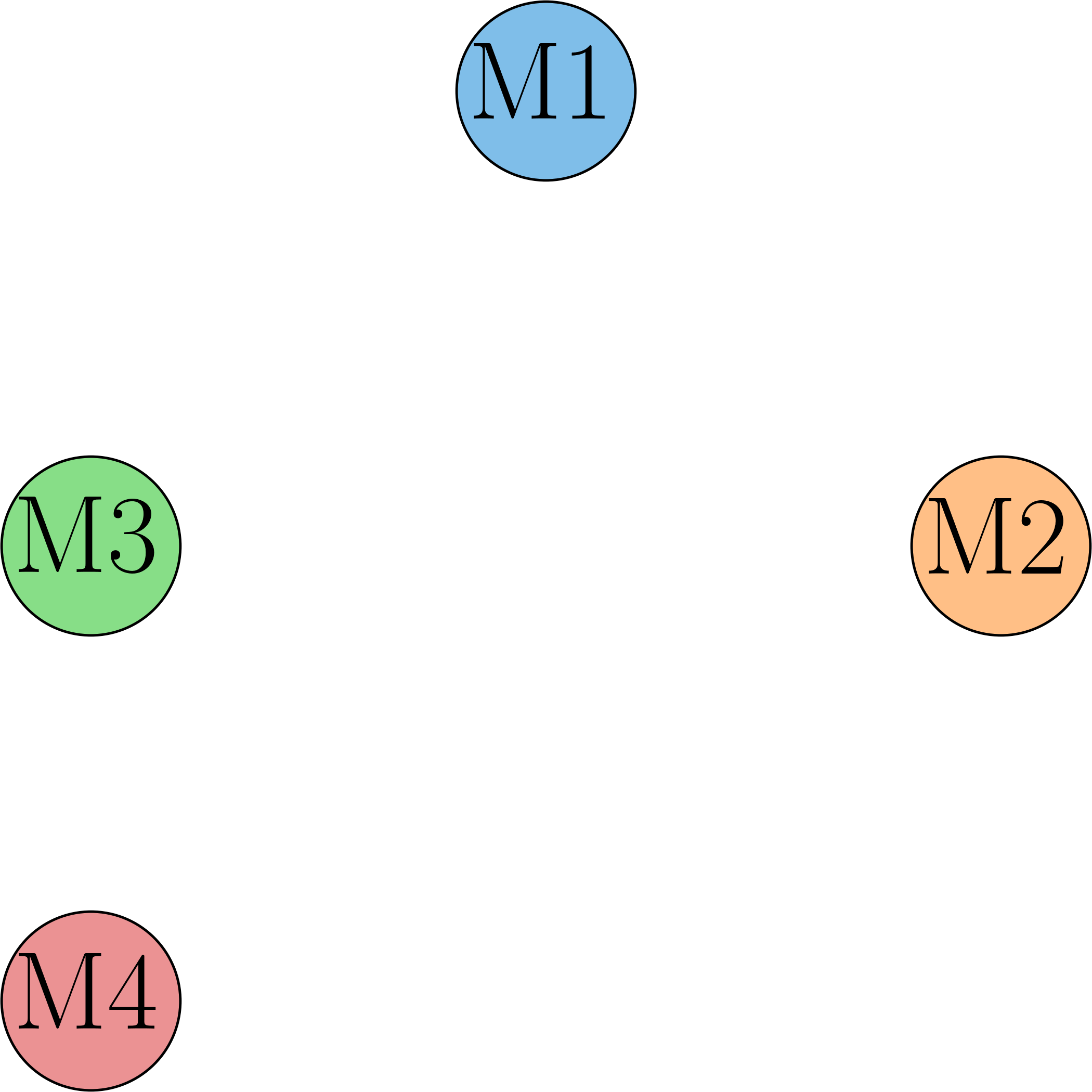}
        \caption{$3$-line graph.}
        \label{fig:meeting_3_line_graph}
    \end{subfigure}
    \hfill
    
    \caption{$s$-line graphs for hypergraph in Fig.~\ref{fig:pipeline_permissible_walks} for $s \in \{0,1,2,3\}$ with undirected edges.}

    \label{fig:sline_graphs_multiple_s}
\end{figure}

For example, given the hypergraph in Fig.~\ref{fig:pipeline_permissible_walks} we can construct its $s$-line graph at various levels of $s$ to capture the strength of the hyperedges connections as shown in Fig.~\ref{fig:sline_graphs_multiple_s}.

\renewcommand{\lg}{{\cal L}_s}

The attributed line graph $\lg$ is derived from the line graph $L_s$ and the attributes of the attributed hypergraph $H = (V, E, \phi, \epsilon, \gamma)$, but has bidirected edges $K$:

\begin{definition}
The \textbf{attributed $s$-line graph} $\mathcal{L}_s = (E,K,\tau,\zeta)$ of an attributed hypergraph $H=(V,E,\epsilon,\phi,\gamma)$ has node set $E$, but now $K \sub E^2$ is a set of directed edges, with both $(e_i,e_{i'}), (e_{i'},e_i) \in K$ if $|e_i \cap e_{i'}| \ge s$, for $e_i \neq e_{i'}$. Here $\tau: E \to A$ is a vertex attribute function to the attribute set $A$ and $\zeta: K \to B$ is an edge attribute function to the edge attribute set $B$.
%Additionally, $\mathcal{L}_s$ is a directed graph where each edge is bidirectional.
\end{definition} 

This definition is quite general. But in this work we simply use $A=Y$ and $\tau = \epsilon$, so that the vertex attributes of the line graph $\mathcal{L}_s$ are simply the  edge attributes of the source hypergraph $H$. Note also that the edge attribute function $\zeta$ of $\lg$ need not have a direct representation in the hypergraph, as it is on {\em ordered pairs} of $s$-incident edges. While available in our formalism, in this paper we will not use $\zeta$ further. 

Our last step is to generate the permissible walk graph from the attributed $s$-line graph by considering relations on the attributes.

\begin{definition} \label{def:permissible_s_line_graph}
For an attributed hypergraph $H=(V,E,\epsilon,\phi,\gamma)$, let $q: A\times A \rightarrow \{0,1\}$ be called a {\bf predicate}. Then given an attributed $s$-line graph $\mathcal{L}_s = (E,K, \tau, \zeta)$, the \textbf{permissible walk graph} $P_q = (E, Q, \tau, \zeta|_{Q})$ for a predicate $q$ is a spanning subgraph of $\mathcal{L}_s$ with edge set $Q \subseteq K$, where $Q$ contains all edges $(e_j,e_j')\in K$ satisfying $q(\tau(e_j),\tau(e_{j'})) = 1$. That is, $Q=\{(e_j, e_{j'}) \in K \colon q(\tau(e_j),\tau(e_{j'})) = 1\}$. 
\end{definition}

Given a predicate $q : A\times A \rightarrow \{0, 1\}$, we can define a directed graph $R_q$ with vertex set $E$ (the same vertex set as the line graph $L_s$), and $(e_j,e_j')$ is an edge of $R_q$ if $q(\epsilon(e_j),\epsilon(e_j'))=1$, i.e. if the edge attributes of $e_j$ and $e_j'$ are mapped to $1$ under $q$.
%must take pairs of attribute data in the domain $A\times A$ and result in a binary evaluation for satisfying the relation. The predicate function as defined here gives us the directed adjacency function of a graph on the vertices of the line graph. 
We will call this graph $R_q$ the \textbf{attribution graph}. One way to understand the permissible walk graph is that the perissible walk graph is the (graph) intersection of $R_q$ and $\mathcal{L}_s$, with vertex and edge atributes inherited from $\mathcal{L}_s$.
%the $s$-line graph and attribution graph both define directed graphs on the edge set of the hypergraph and the permissible walk graph is obtained by taking the intersection of their edge sets. 
However, in practice, it is more computationally efficient to compute the permissible walk graph by computing $Q$ from just the edge set $K$. Additionally, the operations of predicate and graph intersections commute.
%Claim that these operations commute: predicate intersection vs graph intersection}
%Additionally, the permissible walk graph is a generalization of the $s$-line graph if the edge attributes $B$ are the intersection size between hyperedges and the predicate $q$ function thresholds the intersections by $s$ resulting in $P_q = R_q \cup L_s = L_s$.

Notice as well that where the edge set $K$ in the attributed line graph $\lg$ is bidirected in that each pair of vertices is connected by a dual pair of directed edges, one in each direction, after intersection with the attribution graph $R_q$, the pairs of vertices in the permissable walk graph $P_q$ may be connected by directed edges in $Q \sub K$ in either direction, both directions, or not at all.

While relations and predicate functions can be defined on edge, incidence, or vertex attributes, it may be more useful to associate (or transfer) the attributes from one attribute location to another. For example, we might have vertex or edge attribute functions be induced from a given incidence attribute function, which we refer to as vertex (resp. edge) \textbf{marginalization}. For another example, we might use given vertex and edge attribute function to create an induced incidence function, which we refer to as an \textbf{extension}. We demonstrate a marginalization function in our toy example in Section~\ref{ssec:examples}

Above we discussed how we would generally be ignoring the possibility of edge attributes $B$ on the vertices of the attributed line graph $\lg$ (which vertices represent {\em pairs} of incident edges in the source attributed hypergraph $H$). But if there were a relevant edge attribution function $\zeta$ the predicate function could also be defined on the edge attributes where $q: B \to \{0,1\}$. An example of how this construction could be useful is if $\zeta(e_j,e_{j'}) = |e_j \cap e_{j'}|$. In this case $B$ is the natural numbers and if for $k \in B$ we have $q(k) = 1$ if $k\geq s$ and $0$ otherwise we can get the $s$-line graph as a permissible walk graph on the line graph of the hypergraph. This shows that the permissible walk framework is strictly more general than the $s$-line framework as the $s$-line graph can be generated from our permissible walk graph. However, we build all of our permissible walk graphs on top of the $s$-line framework for technical simplicity since it is incorporated into all of our later examples.

The resulting permissible walk graph $P_q$ captures information on both the strength of connectivity between hyperedges through the $s$ parameter as well as specific relations desired between hyperedges. 

%!TEX root = ..\main.tex
%-------------------------------------------------
%*************************************************

\section{Instantiations and Cases} \label{ssec:implementation}
In this section, we provide two examples demonstrating the application of the permissible walk graph as well as several implementation details.

While an attributed hypergraph can have any type of hyperedge attributes, our method requires that a predicate $q$ is defined on the set of vertex attributes $A$.
%to form a permissible walk graph.
Some common attribute types that would allow for a relation to be defined are as follows: 
\begin{itemize}
    \item {\bf {Logical Attributes}}: The attribute functions can take values in the set $\{0,1\}$. Relations and the corresponding predicate function can then be defined based on logical expressions (e.g., and/or).
    \item  {\bf Semantic Type Hierarchy Attributes}: The attributes are elements of a hierarchical data structure,  for example a semantic class hierarchy as used in a computational ontology \cite{UsMGrM96}, which has an underlying poset. The relation can then be implied by the hierarchy.
    \item  {\bf Set Attributes}: For the set attribute type, consider $\phi$, $\epsilon$, and $\gamma$ taking values in a set without any additional structure applied. 
We can define a relation on these sets using an intersection predicate such that they are related if they intersect by at least $t$ elements of the sets. 
We consider this attribute type in our toy example with sets attributed on each incidence as shown in Fig.~\ref{fig:topic_attributes}.
    \item  {\bf Temporal Attributes}: For temporal attributes, the functions $\phi$, $\epsilon$, and $\gamma$ have temporal intervals (start time less than or equal to end time) as the attribute. Relations between intervals can be defined in many ways including weak, strong, or inclusion orders~\cite{joslyn2017interval}. An example of interval attributes on the edges as $\epsilon: E \to Y$ is shown in Fig.~\ref{fig:interval_attributes} of our toy example.
\end{itemize}

A common occurrence in attributed hypergraphs is multiple attribute types existing on the edges (e.g., time and logical attributes). In this case given a relation of interest on each attribute type one can establish a natural relation on all pairs of these attribute types by asking that both relations are satisfied by their respective attributes. This construction yields the same permissible walk graph as constructing one for each attribute and taking their intersection; however, it is notationally burdensome. In light of this in the following examples when there is more than one attribute present the authors opt to create a permissible walk graph for each attribute and then intersect them to obtain a permissible walk graph over both attributes.

\subsection{Toy Example} \label{ssec:examples}
\begin{figure}[h!] 
    \centering
    \begin{subfigure}[b]{.28\textwidth}
        \centering
        \includegraphics[width=.99\textwidth]{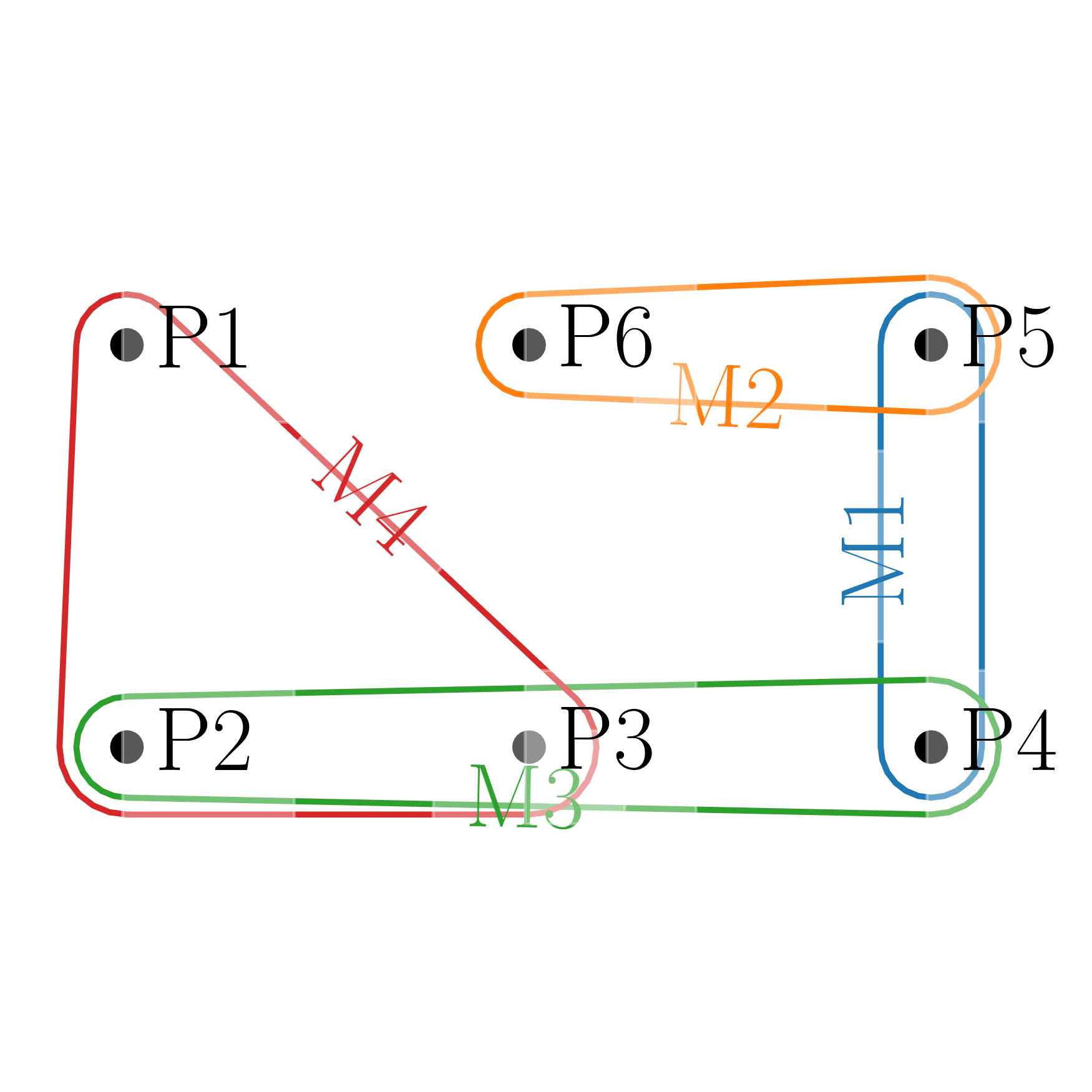}
        \vspace*{-7mm}
        \caption{Hypergraph of meetings.}
        \label{fig:meeting_HG}
    \end{subfigure}
    \hfill
    \begin{subfigure}[b]{.28\textwidth}
        \centering
        \includegraphics[width=.99\textwidth]{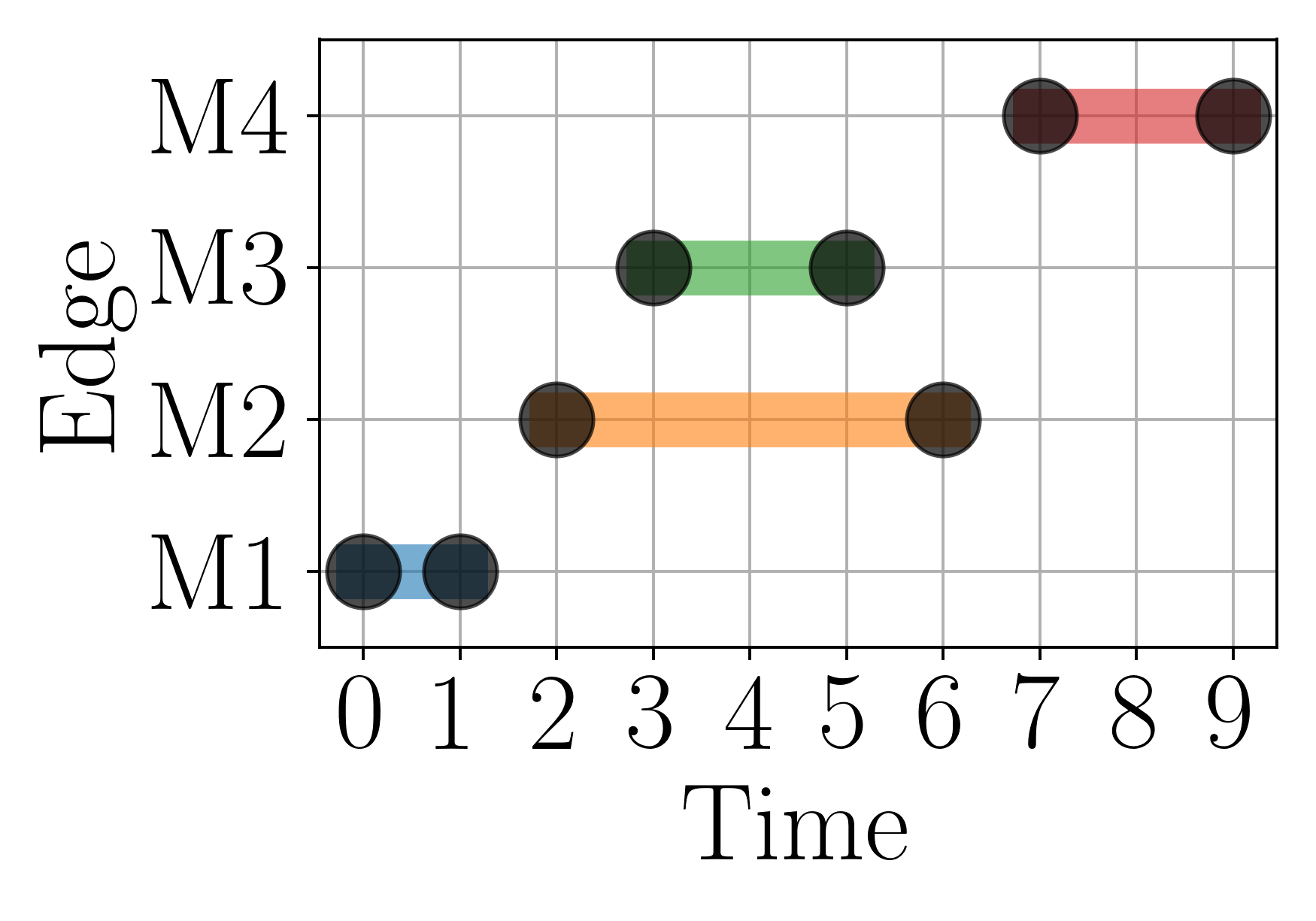}%
        \caption{Time interval edge attributes.}
        \label{fig:interval_attributes}
    \end{subfigure}
    \hfill
    \begin{subtable}[b]{.4\textwidth}
        \centering
        \resizebox{\textwidth}{!}{
        \begin{tabular}{cccccc}
\multicolumn{1}{l}{} & \multicolumn{1}{l}{}           & $\{A, B, C\}$ & $\{E, F\}$  & $\{B, C, D\}$ & $\{C\}$     \\
\multicolumn{1}{l}{} & \multicolumn{1}{c|}{\textbf{}} & $M_1$         & $M_2$       & $M_3$         & $M_4$       \\ \cline{2-6} 
$\{C\}$              & \multicolumn{1}{c|}{$P_1$}     & $\emptyset$   & $\emptyset$ & $\emptyset$   & $\{C\}$     \\
$\{C, D\}$           & \multicolumn{1}{c|}{$P_2$}     & $\emptyset$   & $\emptyset$ & $\{C, D\}$    & $\{C\}$     \\
$\{B, C\}$           & \multicolumn{1}{c|}{$P_3$}     & $\emptyset$   & $\emptyset$ & $\{B, C\}$    & $\{C\}$     \\
$\{A, B, D\}$        & \multicolumn{1}{c|}{$P_4$}     & $\{A, B\}$    & $\emptyset$ & $\{D\}$       & $\emptyset$ \\
$\{A, C, E, F\}$     & \multicolumn{1}{c|}{$P_5$}     & $\{A, C\}$    & $\{E, F\}$  & $\emptyset$   & $\emptyset$ \\
$\{F\}$              & \multicolumn{1}{c|}{$P_6$}     & $\emptyset$   & $\{F\}$     & $\emptyset$   & $\emptyset$
\end{tabular}}
        \vspace*{6mm}
        \caption{Topic incidence attributes.}
        \label{fig:topic_attributes}
    \end{subtable}
    \caption{Attributed hypergraph of meetings with meeting time intervals as edge attributes and topics as incidence attributes.}
    \label{fig:attributed_meeting_HG_example}
\end{figure}

We now build our first example hypergraph, which we use to motivate and demonstrate various aspects and uses of the permissible walk graph. Code for replicating this toy example will be available through the permissible walks module of HyperNetX. This toy example attributed hypergraph in Fig.~\ref{fig:meeting_HG} is constructed to model four meetings, $M_1 \ldots, M_4$, that occurred between six people, $P_1, \ldots, P_6$, with a total of six topics discussed, $A, B, \ldots, F$. Let $\mathcal{A}:=\{A,B,\ldots, F\}$. These meetings consisted of subsets of people which have both topic and temporal attributes. Specifically, we placed interval attributes on the edges as shown in Fig.~\ref{fig:interval_attributes} and topic attributes on the incidences as shown in Fig.~\ref{fig:topic_attributes}. The incidence attributes $Z$ of topics in Fig.~\ref{fig:topic_attributes} represent which topics were discussed by which people in each meeting (e.g., person $P_4$ discussed topics $A$ and $B$ in meeting $M_1$) and the interval attributes in Fig.~\ref{fig:interval_attributes} describe when the meeting occurred (e.g., $M_1$ occurred from time 0 to 1). We note here that $Z=2^\mathcal{A}$.
From these data we could directly try to tell which meetings influence which. To do this we need to account for both the temporal order of the meetings as well as the topics discussed at each meeting. Let us pose the question whether $M_1$ influenced $M_2$. We can see that $M_1$ did not share any of the same topics as discussed in $M_2$, but they did share person $P_5$ and $M_1$ did occur before $M_2$. The permissible walk graph encapsulates all of this information, as we will now demonstrate.

In this toy example we demonstrate edge marginalization. Given the above incidence attribute function $\gamma$ in Fig.~\ref{fig:topic_attributes}, we get an induced edge attribute function $g_Y\colon E \rightarrow Z$ defined as $g_Y(e) = \bigcup_{v \in R_E(e)} \gamma(v,e)$.
That is, $g_Y$ sends an edge $M_i\in E$ to the union of topic sets. Enumerating all evaluations of $g_Y$ on the edges, we get $g_Y(M_1) = \{A, B, C\}$, $g_Y(M_2) = \{E, F\}$, $g_Y(M_3) = \{B, C, D\}$, and $g_Y(M_4) = \{C\}$. We apply this marginalization since, for example, it may be more useful to know the set of topics discussed in a meeting instead of which topics were discussed by which people.
We also have another example of an edge marginalization function in the results of Section~\ref{sec:results} for temporal attributes by taking the convex hull of intervals attributed to incidences.

We now construct the permissible walk graph from the attributed hypergraph in our toy example where we used an edge marginalization $g_Y$. The hyperedges are attributed with both the temporal attributes given by the original edge attribute function $\epsilon$ (with values in the set of time intervals) and the set attributes given by the function $g_Y$ (with values in the subsets of all topics) induced from the original incidence attribute function $\gamma$.

\begin{figure}[h] 
    \centering
    \begin{subfigure}[t]{.22\textwidth}
        \centering
        \includegraphics[width=.99\textwidth]{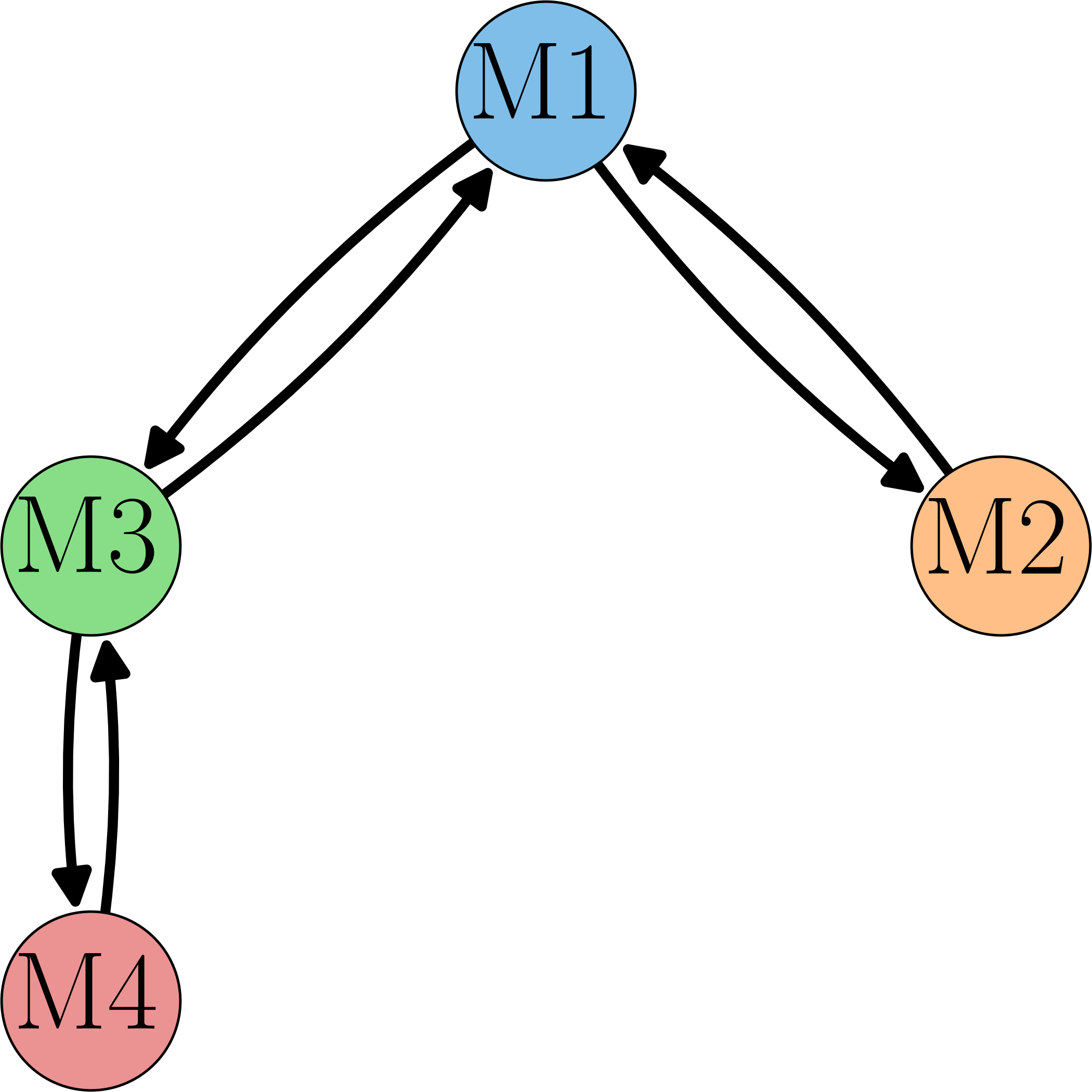}
        \caption{$\mathcal{L}_s(H)$.}
        \label{fig:meeting_s_line_graph}
    \end{subfigure}
    \hspace{5pt}
    \begin{subfigure}[t]{.22\textwidth}
        \centering
        \includegraphics[width=.99\textwidth]{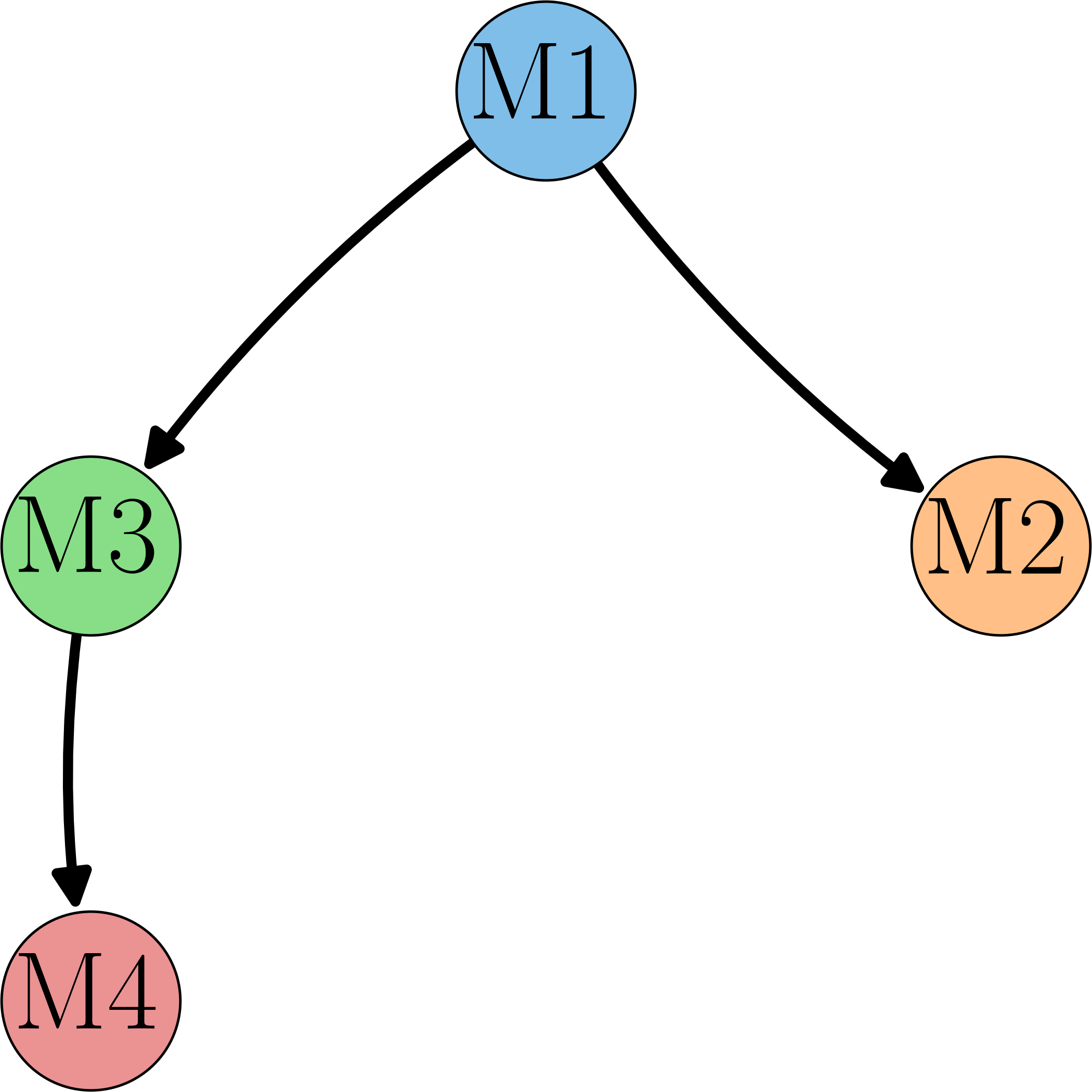}
        \caption{$P_i$.}
        \label{fig:meeting_PW_graph_temporal}
    \end{subfigure}
    \hspace{5pt}
    \begin{subfigure}[t]{.22\textwidth}
        \centering
        \includegraphics[width=.99\textwidth]{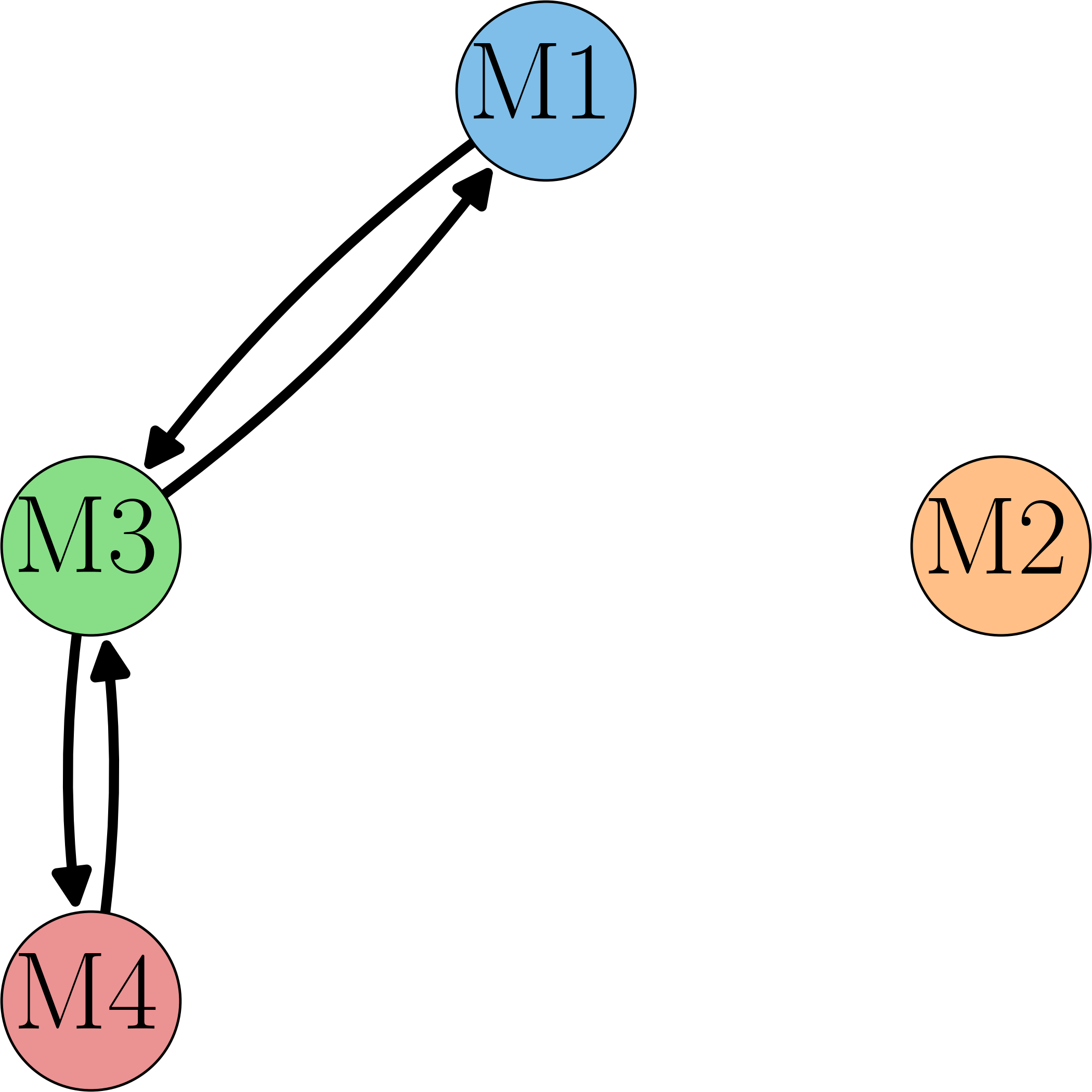}%
        \caption{$P_t$.}
        \label{fig:meeting_PW_graph_relation}
    \end{subfigure}
    \hspace{5pt}
    \begin{subfigure}[t]{.22\textwidth}
        \centering
        \includegraphics[width=.99\textwidth]{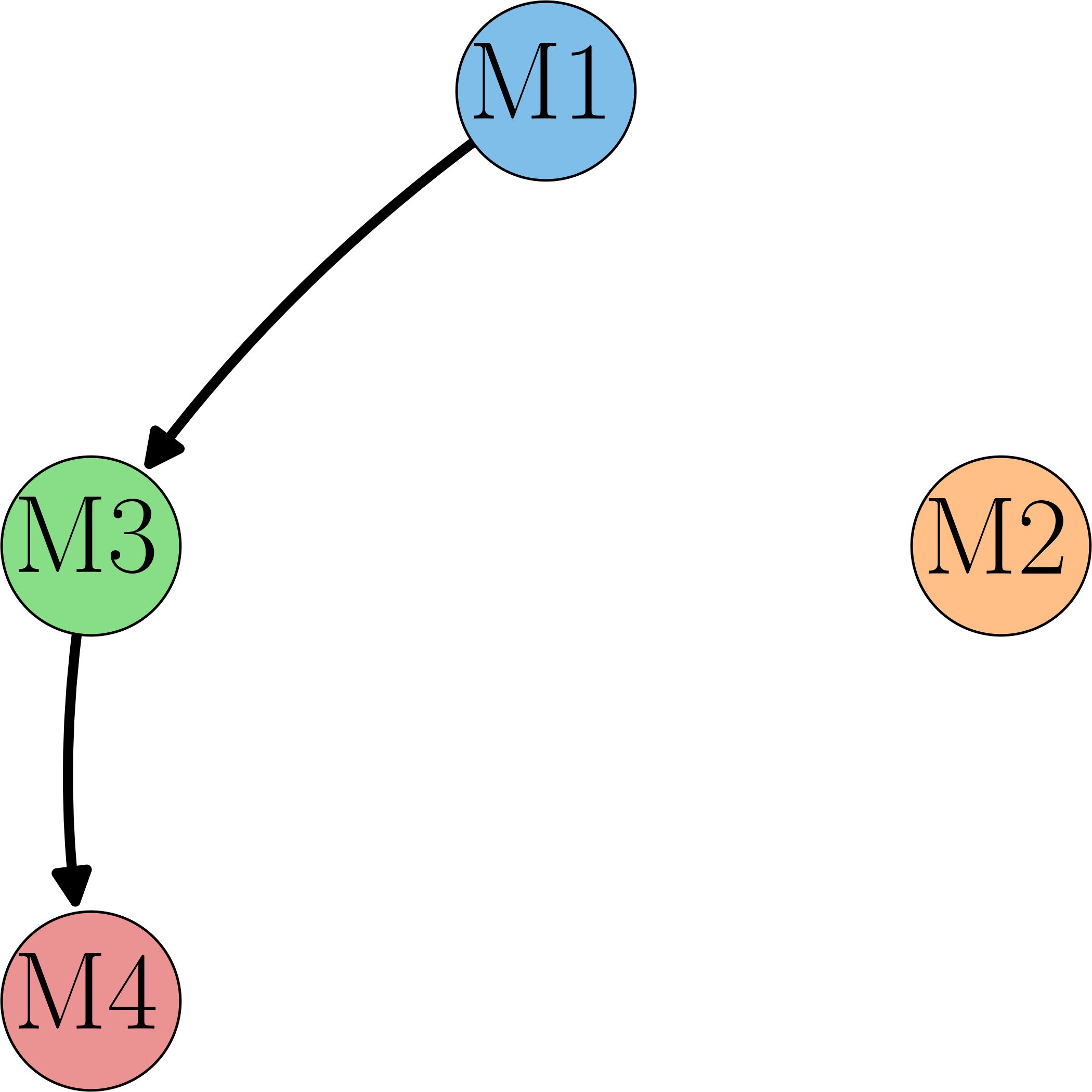}%
        \caption{$P_{it} = P_i \cap P_t$.}
        \label{fig:meeting_PW_graph_relation_temporal}
    \end{subfigure}
    \caption{(a) The attributed $s$-line graph $\lg$ for $s=1$, (b) interval order permissible graph $P_i$, (c) topic respecting permissible walk graph $P_t$, and (d) interval order and topic respecting permissible walk graph $P_{it}$ for the toy example hypergraph in Fig.~\ref{fig:meeting_HG}.}
    \label{fig:meeting_HG_line_graphs}
\end{figure}

We begin by calculating the attributed $s$-line graph $\mathcal{L}_s$ for $s=1$ of the example hypergraph with only the vertex attribute function $\tau = \epsilon$ being used. For a vertex $e\in E$ of $\mathcal{L}_s$, we have two functions that operate on the nodes: $\epsilon(e)$ and $g_Y(e)$, where $\epsilon(e)$ captures the interval attributes and $g_Y(e)$ the marginalized topic attributes.
Fig.~\ref{fig:meeting_s_line_graph} shows the attributed $s$-line graph. Note that each of these permissible graphs are directed specifically in virtue of the corresponding predicate $q$.

With the attributed $s$-line graph we now construct various permissible walk graphs using either or both of the node attributes. To do this we need to define the predicate functions.
The first predicate we use for the interval attributes $i:\mathcal{I}\times \mathcal{I} \rightarrow \{0, 1\}$, where $\mathcal{I}$ is the set of intervals on $\mathbb{R}$, is based on the strong interval order \cite{joslyn2017interval}: 
\begin{equation}   
i([a, b], [c, d]) = 
\begin{cases}
    1, & \text{if } b \leq c \\
    0, & \text{if } b > c,
\end{cases}
\label{eq:strong_order}
\end{equation}
where $[a, b], [b,c] \in \mathcal{I}$.
This function captures one sense of the temporal directionality of the meetings, where if one meeting ends before another starts, there is a flow from that meeting to the next. We construct the corresponding permissible walk graph $P_i$ by including only the edges $Q = \{(e_i, e_j) \in K \colon i(\epsilon(e_i), \epsilon(e_j)) = 1\}$ with $\epsilon$ as the edge attribute function of the attributed hypergraph, see Fig.~\ref{fig:meeting_PW_graph_temporal}.

The second function $t:\mathcal{A} \times \mathcal{A} \rightarrow \{0, 1\}$ only accounts for the topic sets on the vertices and determines if the sets intersect as 
\begin{equation}   
t(S_1, S_2) = 
\begin{cases}
1, & \text{if } S_1 \cap S_2 \neq \varnothing \\
0, & \text{else},
\end{cases}
\label{eq:intersection_topics}
\end{equation}
where $S_1$ and $S_2$ are sets.
This functions checks to see which meetings have related topics, but has bidirectional edges as $t(S_1,S_2) = t(S_2,S_1)$. We construct the permissible walk graph $P_t$ by including only the edges $Q = \{(e_i, e_j) \in K \colon t(g_Y(e_i), g_Y(e_j)) = 1\}$, see Fig.~\ref{fig:meeting_PW_graph_relation}.

The last permissible walk graph $P_{it} = P_i \cap P_t$ captures both the temporal directionality of the meetings as well as how the topics are related by taking the intersection of the other two permissible walk graphs. $P_{it}$ is shown in Fig.~\ref{fig:meeting_PW_graph_relation_temporal}. With $P_{it}$ we are able to now answer questions such as ``did meeting $M_1$ influence the discussion in meeting $M_2$?'' The permissible walk graph indicates that there is no observable mutual influence between $M_1$ and $M_2$ since there is no edge between the two. However, we can observe that $M_1$ could have influenced $M_3$ and $M_4$ as there are edges from $M_1$ to either and yet neither influenced $M_1$ due to the temporal directionality from $i$.

%!TEX root = ..\main.tex
%-------------------------------------------------
%*************************************************

\section{Application: Social Network Analysis} \label{sec:results}

In this section we consider the application of permissible walks to the Reddit social media network, where the vertices are users and the hyperedges represent discussion threads in which the users post messages. These structures are temporally attributed in that there are time stamps on messages, supporting time intervals on users, threads, and users' activity in threads, in terms of the initial and final posting times. 

If we wanted to see the flow of traffic through posts on particular ``subreddits'' (topic boards within Reddit as a whole) we could use $s$-line graphs to recover which posts share a certain number of users. But this would  fail to capture the direction of the flow over time; i.e., users going from one post to another. 
%The want to measure the flow of a certain amount of traffic with respect to time brings us to this analysis. 
Our first analysis studies the intra-subreddit permissible walks by demonstrating how the permissible walk graph can capture the influence of a thread on other threads. 
Following this we study the inter-subreddit relations by showing the interaction matrix between subreddits as a summary of the permissible walk graph, which captures information about connectivity and directionality of communication between subreddits.
Additionally, throughout this analysis we make comparisons to the standard $s$-line graph approach that could be used to study the interactions between subreddits.

\subsection{Data}
The Reddit dataset is from the PAPERCRANE Social Network Problem-shop ~\cite{Papercrane2022}.
The dataset consists of COVID-19 related subreddits for 320 days from January 15 to November 30, 2020. 
The data is of the author posts, which includes information on the time of post, thread ID, and subreddit. 

Over the duration of the study time, $n$ users have posted at least once to at least one of $m$ different threads. As each post is time-stamped, there is an $n \times m$ ``post array'' $W$ whose cells $W[i][j], i \in [n], j \in [m]$ contain a list of the time stamps for the posts that user $i$ made to thread $j$. For each such cell $W[i][j]$, we can then generate a time interval $I_{i,j}$ simply as the min and max extrema:
    \[ 
    I_{i,j} := \left[ \min W[i][j], \max W[i][j] \right]. 
    \]
    Given this array of intervals, we can derive intervals which describe the time frame in which a thread was active, and similarly when a user was active. We define $I_i^{row}$ and $I_j^{col}$ to be the convex hull over intervals in the $i$th row and $j$th column respectively. 
    %the time intervals $I_i$ as row marginalizations on users, and $I_j$ as column marginalizations on threads, are derived by ``union over gaps'' or convex hull (denoted $\uplus$) over the rows and columns respectively, 
    Specifically, let $\uplus$ denote the convex hull, giving us:
    \[ 
    I_i^{row} = \mathop{\uplus}\limits_{j=1}^m I_{i,j} = \left[ \min_{j=1}^m I_{i,j}, \max_{j=1}^m      I_{i,j} \right], \qquad
    I_j^{col} = \mathop{\uplus}\limits_{i=1}^n I_{i,j} = \left[ \min_{i=1}^n I_{i,j}, \max_{i=1}^n         I_{i,j} \right]. 
    \]
    We note that $I_i^{row}$ is a row marginalization on users, and $I_j^{col}$ is a column marginalization on threads.
    This is cartooned in Table ~\ref{fig:post_array_reddit}, showing the user and thread time intervals on the margins of the post array. In the example, $I_{n,1}=[5,10]$, while $I_{j=m}^{col} = [13,13] \uplus [2,7] = [2,13]$. Note that user 1 did not post to thread $m$, nor did users 2 or $n$ post to thread 2.

\begin{table}[h!]
        \centering
        \caption{The post array $A$ with a list of post times in the cells and marginal time intervals on users and threads.}
        \begin{tabular}{cccccc}
        \multicolumn{1}{l}{} & \multicolumn{1}{l}{}           & $[1, 10]$ & $[4, 7]$  &  & $[2,13]$     \\
        \multicolumn{1}{l}{} & \multicolumn{1}{c|}{\textbf{}} & Thread $1$  & Thread $2$  & $\cdots$  & Thread $m$  \\ \cline{2-6} 
        
        $[1,7]$   & \multicolumn{1}{c|}{User $1$}  & $1, 2, 3$ & $4, 7$ & $\cdots$ & $\emptyset$ \\
        
        $[8,13]$  & \multicolumn{1}{c|}{User $2$}  & $8$ & $\emptyset$ & $\cdots$ & $13$ \\
        
        $ $       & \multicolumn{1}{c|}{$\vdots$}  & $\vdots$ & $\vdots$ & $\cdots$ & $\vdots$ \\
        
        $[2,10]$  & \multicolumn{1}{c|}{User $n$}  & $5, 10$ & $\emptyset$ & $\cdots$ & $2, 7$ \\
        \end{tabular}
\vspace*{6mm}
\label{fig:post_array_reddit}
\end{table}

\begin{figure}[h!] 
    \centering
    \includegraphics[width=.99\textwidth]{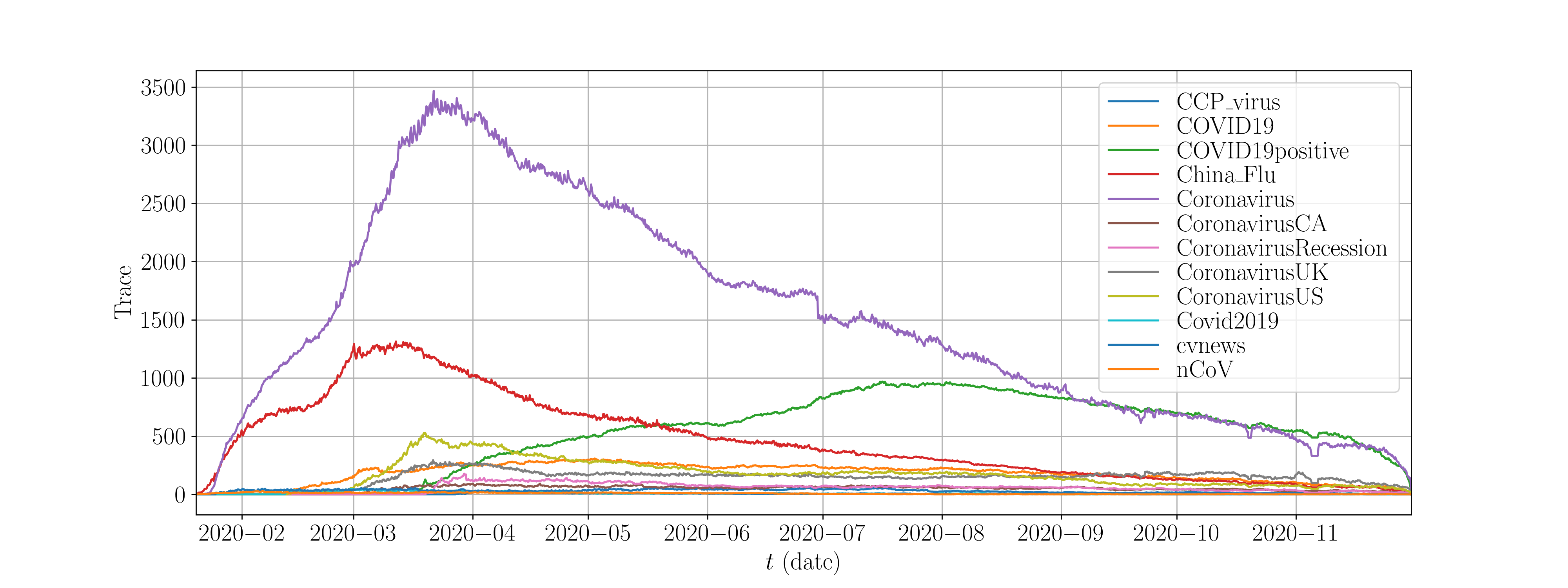}
    \caption{Trace  $T(t)$ (interval count at time $t$) of the collection of temporal intervals $I_j$ associated to each thread for COVID-19 related subreddits from the PAPERCRANE~\cite{Papercrane2022} dataset.}
    \label{fig:trace_of_reddit_data_per_subreddit}
\end{figure}

The dataset includes 12 subreddits which are listed in Fig.~\ref{fig:trace_of_reddit_data_per_subreddit}. This figure also helps visualize the size of each subreddit over the 10 months of data by showing the trace of intervals associated to the threads. 

\newcommand{\I}{{\cal I}}

\begin{definition}
Assume there exists a collection of intervals ${\cal I}=\{ I_i \}_{i=1}^n$ over support $U(\I) = \mathop{\uplus}\limits_{i=1}^n I_i$, then their \textbf{trace} \cite{JoCBoJ05} is a function $\func{T}{U(\I)}{\ints_{\ge 0}}$ 
%of a collection of intervals $I$ and time instances $t = [t_0, t_1, \ldots, t_n]$ 
which counts the number of active intervals at any time $t \in U(\I)$, so that 
    \[ T(t) = |\{I_i \in \I | \min(I_i) \leq t \leq \max(I_i)\}|. \]
\end{definition}
\fig{fig:trace_of_reddit_data_per_subreddit} shows the traces of the thread intervals $I_j$ for the subreddits, with $t$ limited to 2000 evenly spaced instances spanning the support of the interval collection. 
%The closed intervals were defined for each thread as the time from the first to last post of that thread.

\subsection{Intra-Subreddit Analysis}

The first motivation for studying the PAPERCRANE dataset using permissible walks is to gain some understanding of the intra-subreddit communication structure: which threads within a subreddit influence other threads? 

To build this intra-subreddit analysis, we will use an example for the subreddits China\_Flu and Coronavirus in January 2020. During these first two weeks of the data, both subreddits are growing in popularity as shown in Fig.~\ref{fig:trace_of_reddit_data_per_subreddit}. Our goal is to see how the permissible walk graphs evolve during these two weeks and the spread of influence of an example thread on all the other threads.

\begin{figure}[h!] 
    \begin{center}
    \includegraphics[width=.99\textwidth]{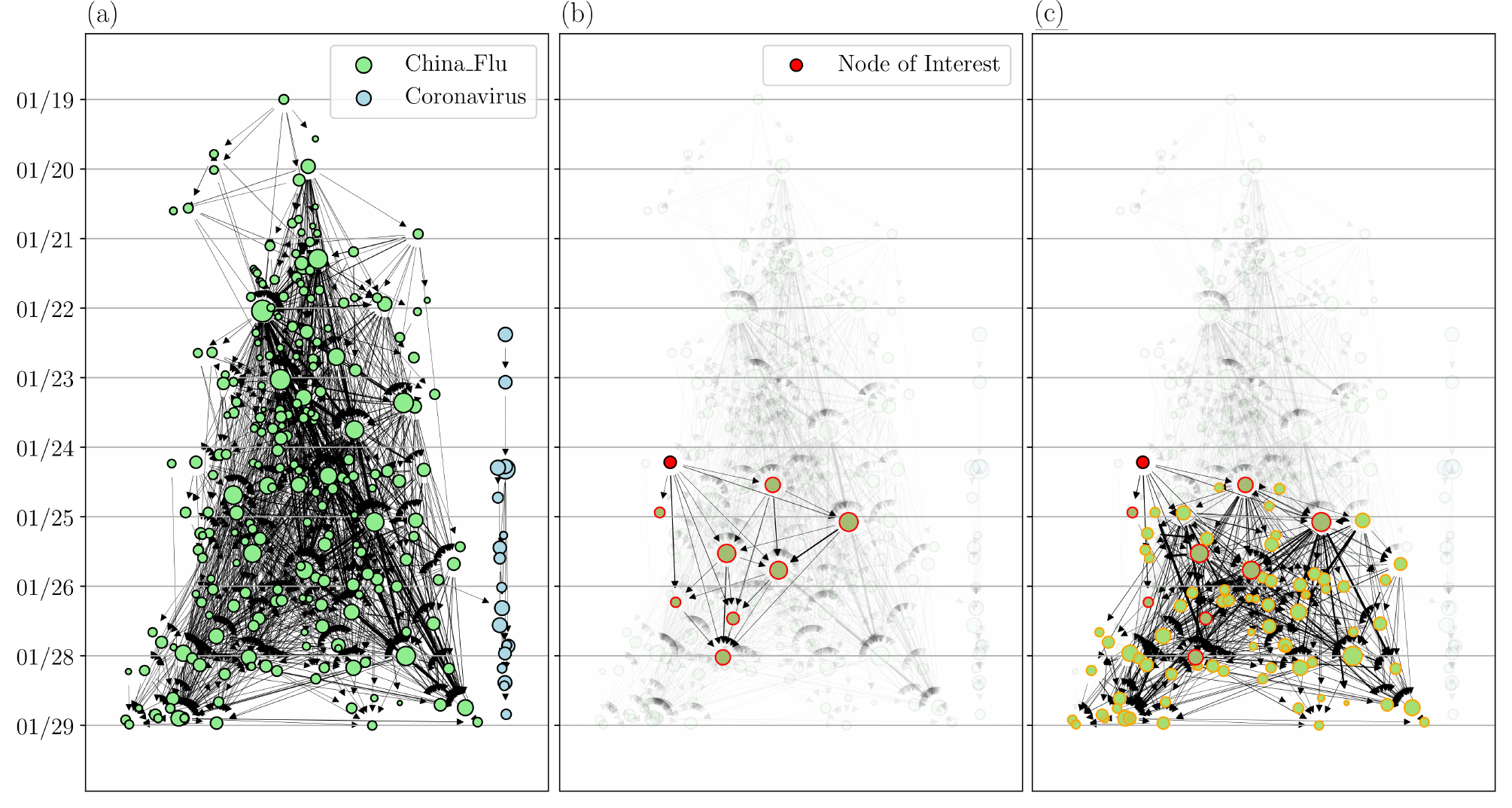}
    \end{center}
    \caption{Intra-subreddit analysis for Coronavirus and China\_Flu subreddits using the permissible walk graph (a) as well as the directly connected (neighbor) downstream nodes of a node of interest (b) and all downstream nodes of the node of interest (c).}
    \label{fig:intra_subreddit_analysis}
\end{figure}

Figure~\ref{fig:intra_subreddit_analysis} shows the color coded permissible walk graph for the China\_Flu and Coronavirus subreddits. This graph was constructed with $s=10$ to allow for visualization as the graph is too large to practically display for lower $s$. Additionally, we used the strong order predicate as described in Section~\ref{sec:method}-Eq.~\eqref{eq:strong_order}. For visualization we have also removed any isolated nodes (85 for this example) in the graph. Additionally, the node sizes and edge widths are correlated to the number of authors in a thread and the number of intersecting authors between threads, respectively. 

We see that initially, China\_Flu is the dominant subreddit. However, around January 22 Coronavirus begins to have threads with size $\geq 10$. Additionally, the graph for this $s$ shows that there are clearly two main communities that are separated by subreddit assignment with only a total of 6 edges connecting the two.

To understand the influence of a thread on all downstream threads we used a demonstrative thread of interest (Node of Interest in Fig.~\ref{fig:intra_subreddit_analysis}~(b)). We can see that for this node of interest there are only 8 immediately downstream neighbors, which means that these corresponding threads shared at least 10 coauthors and happened temporally after the node of interest. This provides a visualization and understanding of the direct influence this thread (node) of interest had on these other threads. In Fig.~\ref{fig:intra_subreddit_analysis}~(c) we also show all downstream nodes from the thread of interest which captures the possible range of influence of this node for $s=10$, which may be of importance when answering questions about the possible social effects of the discussion in the node of interest.

A similar analysis would not be possible with just the $s$-line graph as it does not include directionality information. As such, a similar analysis would result in all connected nodes being influenced from the node of interest, which is clearly not possible due to some threads ending before the thread of interest began.

\subsection{Inter-Subreddit Analysis}

Additionally, this analysis can also be done for inter-subreddit analysis for understanding the exact threads of interaction between subreddits.
%We now want to understand the influence of subreddits on other subreddits. 
To do this, we utilize the interaction matrix $M$. For our purposes, we will consider each subreddit a ``class''. 
\begin{definition}
Let $\mathcal{C}=\{c_1,\dots,c_r\}$ be a set of classes, and let $G$ be a node attributed directed graph with class type as an attribute, i.e. we have a vertex attribute function $C:V(G)\to \mathcal{C}$.
%where the class $c$ is retrieved with an attribute function $c = C(n)$ for each node $n$. 
We define the \textbf{interaction matrix} $M$ to be the $r\times r$ matrix where $M_{i,j}$ is given by the number of edges $(s,t)\in E(G)$ such that $C(s)=c_i$ and $C(t)=c_j$.
%as a weighted, directed adjacency matrix describing the class interaction graph with class nodes.
%For each edge $e = (s, t) \in E(G)$, $M_{i, j}$ is incremented if $C(s) = c_i$ and $C(t) = c_j$.
\end{definition}

\begin{figure}[h!] 
        \centering
        \includegraphics[width=\textwidth]{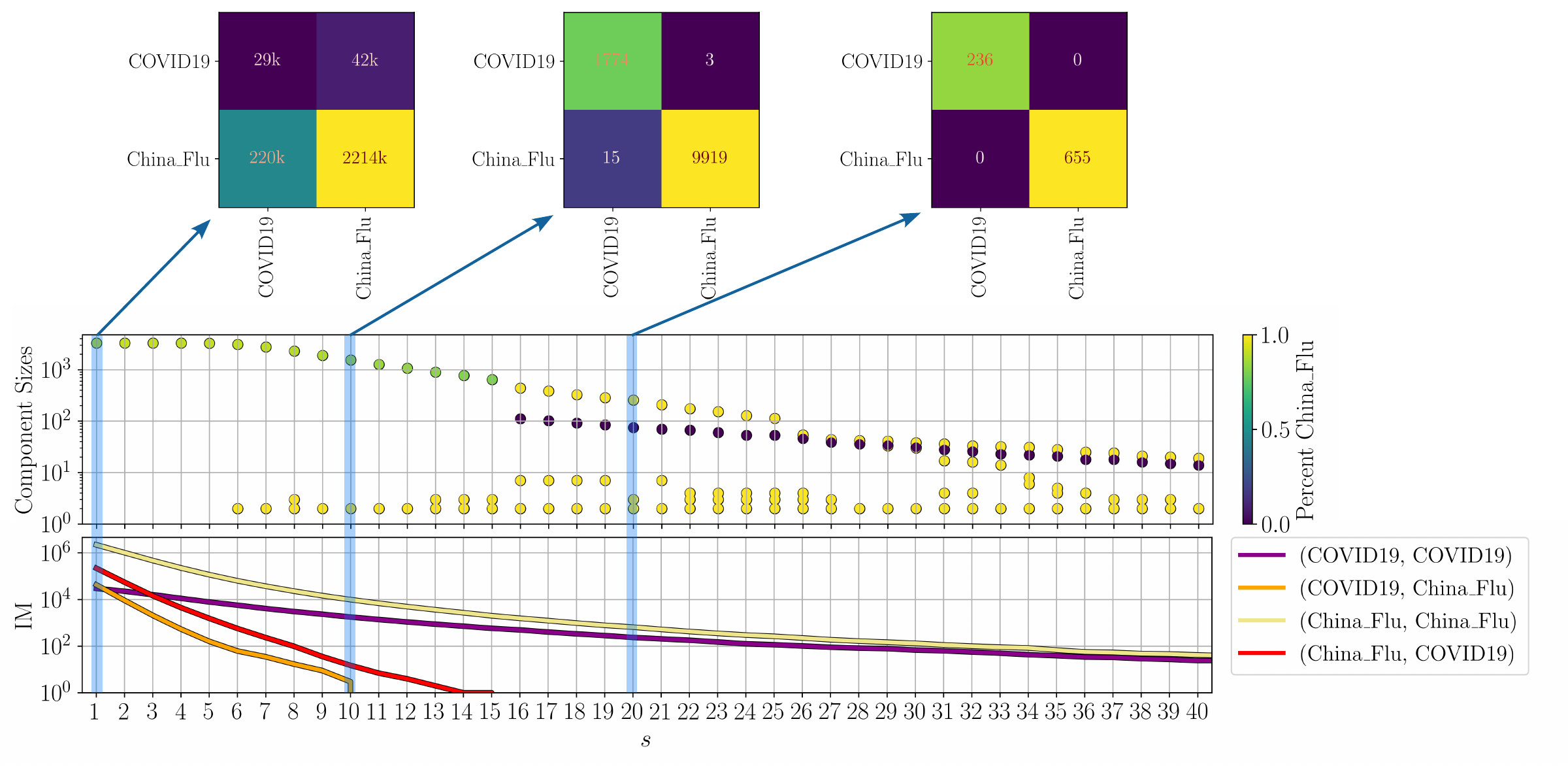}
        \caption{Interaction matrices at various $s$-values demonstrating connectivity of edges between subreddits COVID19 and China\_Flu in the permissible walk graph. Additionally, the number of components and their relative size are provided. In the top chart, each point represents the number of nodes in a weakly-connected component in the permissible walk graph. In the bottom figure are the values of the interaction matrices as $s$ increases showing the changing connectivity between the two subreddits in the permissible walk graph.}
        \label{fig:IM_china_flu_covid19_scaling_s}
\end{figure}

We can think of the interaction matrix as the adjacency matrix of a weighted, directed graph (which may contain directed loops) on the node set $\mathcal{C}$ which describes the interactions between classes in $G$, which we will call the \textbf{class graph}. For the Reddit data, our classes are the subreddits which the given threads are a part of, so the class graph has subreddits as nodes and has weighted, directed edges based on the number of edges in the permissible walk graph from one subreddit to another.
It should be noted that the interaction matrix does not capture information about how many components there are within a subreddit, just the number of connecting edges between subreddits. However, as is shown in Fig.~\ref{fig:IM_china_flu_covid19_scaling_s}, we include the number of components and show that the subreddits tend to predominantly be composed on a single component.
This formation captures the inter-subreddit communication. If the resulting interaction matrix is zero off-diagonal, then there is no interaction between subreddits.

We now demonstrate how the interaction matrix captures information about both connectivity of subreddits as well as directionality in the flow of communication between the subreddits of COVID19 and China\_Flu.
For these subreddits there is a high asymmetry in the interaction matrix. As shown in Fig.~\ref{fig:trace_of_reddit_data_per_subreddit}, the subreddits China\_Flu and COVID19 begin at different times. This is due to COVID-19 not yet being official named. Because of this, alternate names such as China\_Flu were used. On February 11  the World Health Organization announced COVID-19 as the official name of coronavirus and consequently, the COVID19 subreddit quickly grew in popularity. We theorize this corresponded to a shift of authors from China\_Flu to COVID19.
This possible influx of authors from China\_Flu to COVID19 is captured using the interaction matrix as shown in Fig.~\ref{fig:IM_china_flu_covid19_scaling_s} for $s=1, 10, 20$.

At $s=1$, there are approximately 220k edges from China\_Flu to COVID19 and only 42k edges from COVID19 to China\_Flu. This shows the asymmetry of authors originally posting in China\_Flu moving to COVID19. Additionally, there are only 29k self edges for COVID19 at $s=1$, which is relatively low compared to the number of edges connecting to China\_Flu.
However, at $s=10$, this switches from with 1.7k self edges for COVID19, and only a total of 18 edges connecting COVID19 to China\_Flu. This shows that the edges that are connecting the two subreddits do not represent a large number of authors in the intersection of threads. This trend is more clearly shown in the bottom subfigure of Fig.~\ref{fig:IM_china_flu_covid19_scaling_s}, where the number of cross subreddit edges is decreasing at a much higher rate than the self edges.
By increasing $s$ and doing this analysis we are able to show that ``strength" of the connectivity between subreddits and that the threads within a subreddit share more common authors on average per intersection of threads than do cross subreddit threads.
Starting at $s=15$ and shown for $s=20$, we see a bifurcation in the number of components going from one main component to two for each of the subreddits (color coded with yellow as China\_Flu and Purple as COVID19). 
We also note that there are many small components beginning at $s=6$.

%!TEX root = ..\main.tex
%-------------------------------------------------
%*************************************************
\section*{Conclusion} \label{sec:conclusions}

In this work we developed a framework for studying attributed hypergraphs using the permissible walk graph. The permissible walk graph is a generalization of the commonly used $s$-line graph by accounting for hypergraph attributes when creating a directed version of the s-line graph that respects the desired relation between attributes.
The main application we focused on in this work was temporal attributes and how they can be used to improve the analysis of social network data in comparison to the $s$-line graph. However, the permissible walk graph can be used to study many different attribute types that may exist on hypergraphs such as types, hierarchical, sets, and Boolean attributes.

Our results showed that the permissible walk graph was able to capture the spread of communication between threads in multiple COVID-19 related subreddits from the PAPERCRANE reddit dataset~\cite{Papercrane2022}. Additionally, it was able capture the dynamics of authors posting in one subreddit to another by using an interaction matrix as a summary of the permissible walk graph, which is not captured when using the $s$-line graph. Lastly, we showed how the permissible walk graph at different values of $s$ can capture the strength of communication between subreddits with edges connecting subreddits sharing smaller intersections of authors than those within a single subreddit. More generally we hope that this will serve as an example driven explanation of how to use our framework for tracking potential spread of information or disease in attributed higher order networks.

Future research directions using this tool include using the permissible walk graph for clustering 
%with comparisons to state-of-the-art methods 
and implementation of graph neural networks for classification. Additionally, we plan to develop new relations for directionality information of ontological attributes to help in the study of knowledge hypergraphs. On the more theoretical side one question that still remains is how to carefully handle marginalization with different attribution types.

\bibliographystyle{siamplain}
\bibliography{references}

\begin{thebibliography}{10}

\bibitem{aksoy2020hypernetwork}
{\sc S.~G. Aksoy, C.~Joslyn, C.~O. Marrero, B.~Praggastis, and E.~Purvine},
  {\em Hypernetwork science via high-order hypergraph walks}, EPJ Data Science,
  9 (2020), p.~16.

\bibitem{araniti2015contact}
{\sc G.~Araniti, N.~Bezirgiannidis, E.~Birrane, I.~Bisio, S.~Burleigh,
  C.~Caini, M.~Feldmann, M.~Marchese, J.~Segui, and K.~Suzuki}, {\em Contact
  graph routing in dtn space networks: overview, enhancements and performance},
  IEEE Communications Magazine, 53 (2015), pp.~38--46.

\bibitem{banerjee2021spectrum}
{\sc A.~Banerjee}, {\em On the spectrum of hypergraphs}, Linear algebra and its
  applications, 614 (2021), pp.~82--110.

\bibitem{battistella2023permissible}
{\sc E.~Battistella, S.~English, R.~Green, C.~Joslyn, E.~Lagoda, V.~Magnan,
  A.~Myers, E.~Nash, and M.~Robinson}, {\em Permissible walks in attributed
  hypergraphs}, in AMS Special Session on Models and Methods for Sparse (Hyper)
  Network Science (a Mathematics Research Communities session), 2023,
  \url{https://meetings.ams.org/math/jmm2023/meetingapp.cgi/Paper/22323}.
\newblock Presented at JMM 2023.

\bibitem{Papercrane2022}
{\sc J.~Baumgartner, S.~Zannettou, B.~Keegan, M.~Squire, and J.~Blackburn},
  {\em The pushshift reddit dataset}.
\newblock PUSHSHIFT, jan 2020,
  \url{https://doi.org/https://doi.org/10.5281/zenodo.3608135}.
\newblock reddit-hazelnut prepared for the Social Network ProblemShop (Jan
  24-Feb 4, 2022). Ottawa, Canada. Derivative of Reddit data obtained via
  pushshift.io API for the period January 1, 2019 to February 28, 2019.

\bibitem{bernardoni2023algebraic}
{\sc W.~Bernardoni, R.~Cardona, J.~Cleveland, J.~Curry, R.~Green, B.~Heller,
  A.~Hylton, T.~Lam, and R.~Kassouf-Short}, {\em Algebraic and geometric models
  for space networking}, arXiv preprint arXiv:2304.01150,  (2023).

\bibitem{carletti2020random}
{\sc T.~Carletti, F.~Battiston, G.~Cencetti, and D.~Fanelli}, {\em Random walks
  on hypergraphs}, Physical review E, 101 (2020), p.~022308.

\bibitem{carletti2021random}
{\sc T.~Carletti, D.~Fanelli, and R.~Lambiotte}, {\em Random walks and
  community detection in hypergraphs}, Journal of Physics: Complexity, 2
  (2021), p.~015011.

\bibitem{chitra2019random}
{\sc U.~Chitra and B.~Raphael}, {\em Random walks on hypergraphs with
  edge-dependent vertex weights}, in International conference on machine
  learning, PMLR, 2019, pp.~1172--1181.

\bibitem{cooper2012spectra}
{\sc J.~Cooper and A.~Dutle}, {\em Spectra of uniform hypergraphs}, Linear
  Algebra and its applications, 436 (2012), pp.~3268--3292.

\bibitem{DavidBoyce2012}
{\sc B.~R. David~Boyce}, {\em Modeling Dynamic Transportation Networks},
  Springer Berlin Heidelberg, 2012.

\bibitem{failla2023attributed}
{\sc A.~Failla, S.~Citraro, and G.~Rossetti}, {\em Attributed stream
  hypergraphs: temporal modeling of node-attributed high-order interactions},
  Applied Network Science, 8 (2023), pp.~1--19.

\bibitem{fraire2021routing}
{\sc J.~A. Fraire, O.~De~Jonck{\`e}re, and S.~C. Burleigh}, {\em Routing in the
  space internet: A contact graph routing tutorial}, Journal of Network and
  Computer Applications, 174 (2021), p.~102884.

\bibitem{grover2016node2vec}
{\sc A.~Grover and J.~Leskovec}, {\em node2vec: Scalable feature learning for
  networks}, in Proceedings of the 22nd ACM SIGKDD international conference on
  Knowledge discovery and data mining, 2016, pp.~855--864.

\bibitem{hayashi2020hypergraph}
{\sc K.~Hayashi, S.~G. Aksoy, C.~H. Park, and H.~Park}, {\em Hypergraph random
  walks, laplacians, and clustering}, in Proceedings of the 29th acm
  international conference on information \& knowledge management, 2020,
  pp.~495--504.

\bibitem{HOLME201297}
{\sc P.~Holme and J.~Saramäki}, {\em Temporal networks}, Physics Reports, 519
  (2012), pp.~97--125,
  \url{https://doi.org/https://doi.org/10.1016/j.physrep.2012.03.001},
  \url{https://www.sciencedirect.com/science/article/pii/S0370157312000841}.
\newblock Temporal Networks.

\bibitem{Husein2019}
{\sc I.~Husein, H.~Mawengkang, S.~Suwilo, and Mardiningsih}, {\em Modeling the
  transmission of infectious disease in a dynamic network}, Journal of Physics:
  Conference Series, 1255 (2019), p.~012052,
  \url{https://doi.org/10.1088/1742-6596/1255/1/012052}.

\bibitem{joslyn2017interval}
{\sc C.~Joslyn, A.~Pogel, and E.~Purvine}, {\em Interval-valued rank in finite
  ordered sets}, Order, 34 (2017), pp.~491--512.

\bibitem{JoCBoJ05}
{\sc C.~A. Joslyn and J.~Booker}, {\em Generalized information theory for
  engineering modeling and simulation}, in Engineering Design Reliability
  Handbook, E.~N. et~al., ed., CRC Press, 2005, pp.~9:1--40,
  \url{https://www.taylorfrancis.com/books/e/9780429204616/chapters/10.1201/9780203483930-14}.
\newblock
  https://www.taylorfrancis.com/books/e/9780429204616/chapters/10.1201/9780203483930-14.

\bibitem{nguyen2018dynamic}
{\sc G.~H. Nguyen, J.~B. Lee, R.~A. Rossi, N.~K. Ahmed, E.~Koh, and S.~Kim},
  {\em Dynamic network embeddings: From random walks to temporal random walks},
  in 2018 IEEE International Conference on Big Data (Big Data), IEEE, 2018,
  pp.~1085--1092.

\bibitem{ostroski2021dynamic}
{\sc M.~Ostroski}, {\em Dynamic graph edge clustering}, in GraphEx, 2021,
  \url{https://graphex.mit.edu/sites/default/files/images/GraphEx-2021-Poster-Ostroski.pdf}.

\bibitem{ostroski2023scalable}
{\sc M.~Ostroski, G.~Sanders, T.~Steil, and R.~Pearce}, {\em Scalable edge
  clustering of dynamic graphs via weighted line graphs}, arXiv preprint
  arXiv:2311.10337,  (2023).

\bibitem{ostroski2021dynamic2}
{\sc M.~A. Ostroski}, {\em Dynamic graph edge clustering: The art of
  conversation... mining}, in Joint Mathematics Meetings, 2021.
\newblock Presented at JMM 2021.

\bibitem{Skyrms2000}
{\sc B.~Skyrms and R.~Pemantle}, {\em A dynamic model of social network
  formation}, Proceedings of the National Academy of Sciences, 97 (2000),
  pp.~9340--9346, \url{https://doi.org/10.1073/pnas.97.16.9340}.

\bibitem{starnini2012random}
{\sc M.~Starnini, A.~Baronchelli, A.~Barrat, and R.~Pastor-Satorras}, {\em
  Random walks on temporal networks}, Physical Review E, 85 (2012), p.~056115.

\bibitem{UsMGrM96}
{\sc M.~Uschold and M.~Gruninger}, {\em Ontologies: Principles, methods and
  applications}, Knowledge Engineering Review, 11:2 (1996), pp.~93--136.

\end{thebibliography}

\section{Acknowledgements}
\label{sec:ack}

This research began at the 2022 American Mathematical Society (AMS) Mathematics Research Communities (MRC) workshop on Models and Methods for Sparse (Hyper)Network Science. We are grateful to this organization for facilitating collaboration among data scientists and mathematicians on problems relevant to practical applications.

Pacific Northwest National Laboratory is a multiprogram
national laboratory operated for the US Department of Energy (DOE) by Battelle
Memorial Institute under Contract No. DE-AC05-76RL01830. 
PNNL Information Release PNNL-SA-198113.

This work represents the views of the authors and is not to be regarded as representing the opinions of the Center for Naval Analyses or any of its sponsors.

%!TEX root = ..\main.tex
%-------------------------------------------------
%*************************************************

\appendix
\addcontentsline{toc}{section}{Appendices}
\section*{Appendices}

\section{Dynamic Multi-Digraphs} \label{app:ostroski}

Another significant temporal graph structure which can be expressed in our formalism is the dynamic multi-digraph \cite{ostroski2021dynamic,ostroski2021dynamic2}.
\begin{definition}
A \textbf{dynamic multi-digraph} $D = (V, E, T)$ with timestamps where the edges are active is a triple $(V, E, T)$, where $V$ and $E$ are sets of nodes and arcs, respectively, and $T: E \to \mathbb{R}_{+}$ is a function that maps each arc $e$ to a timestamp $t$. The timestamp $t$ indicates when the arc $e$ was active.
\end{definition}

Dynamic multi-digraphs have applications in dynamic dyadic relations including social network analysis~\cite{ostroski2021dynamic,ostroski2021dynamic2} where communications are directed (e.g., instant messaging) and similar objects present themselves in space network routing through delay tolerant networks~\cite{fraire2021routing,araniti2015contact} where nodes represent devices in the network and edges are the contacts where one devices is able to communicate to another at a specific moment in time.

In \cite{ostroski2021dynamic,ostroski2021dynamic2} dynamic multigraphs are studied by using a permissible subgraph of the corresponding line graph referred to as the \textit{increment-weighted line graph} and then applying a clustering algorithm using its minimum spanning tree. Here we can also represent the line graph using our framework by considering the $2$-uniform (edges with only dimension $2$) hypergraph. 
We store the directionality as a hyperedge attribute using a tuple with directed edge $e_i = (n_s, n_t)$  as $\epsilon_d(e_i) = (n_s, n_t)$ with projection functions $\epsilon_d^s(e_i) = n_s$ and $\epsilon_d^t(e_i) = n_t$. 
We additionally have the dynamic time stamps attributed on the hyperedges as $\epsilon_t(e_i) = t$ given edge $e_i$ was active at time $t$. 
To construct a permissible walk graph that satisfies both of the attributes on the edges we first construct two different different permissible walk graphs: the first with $\epsilon = \epsilon_d$ and predicate function $\epsilon_d^t(e_i) = \epsilon_d^s(e_j)$ and the second with $\epsilon = \epsilon_t$ with predicate function $\epsilon_t(e_i) \leq \epsilon_t(e_j)$. 
Here the predicate function is applied for each directed edge $k = (e_i, e_j) \in K$ of the attributed $s$-line. 
The intersection of these two permissible walk graphs gives us the permissible walk graph that satisfies both the temporal order and the directionality of the edges.
By constructing the permissible walk graph in this way on a dynamic multi-digraph we create the same line graph that was used in~\cite{ostroski2021dynamic,ostroski2021dynamic2} which shows how our method is a generalization allowing for future similar applications but with multi-way interactions native to hypergraphs instead of dyadic interactions from graphs.

\end{document}